\documentclass[runningheads]{llncs}
\usepackage[T1]{fontenc}
\usepackage{graphicx}
\usepackage{tikz}
\usepackage{pgfplots}
\usepackage{pgfplotstable}

\usepackage{float}
\usepackage{multicol}
\usepackage{multirow}
\usepackage{xcolor}
\usepackage{caption}
\usepackage{subcaption}
\usepackage{placeins}
\usepackage[ruled,vlined]{algorithm2e}
\usepackage{amsmath,amsfonts}
\usepackage{algpseudocode}
\usepackage{comment}

\usepackage{array}
\newcolumntype{L}[1]{>{\raggedright\let\newline\\\arraybackslash\hspace{0pt}}m{#1}}
\newcolumntype{C}[1]{>{\centering\let\newline\\\arraybackslash\hspace{0pt}}m{#1}}
\newcolumntype{R}[1]{>{\raggedleft\let\newline\\\arraybackslash\hspace{0pt}}m{#1}}

\newtheorem{assumption}{Assumption}

\begin{document}
%
%\title{Private and Secure Fuzzy Name Matching}
\title{Privacy-preserving Fuzzy Name Matching for Sharing Financial Intelligence}
%\begin{comment}
\author{Harsh Kasyap\inst{1,2} \and
Ugur Ilker Atmaca\inst{1,2} \and
Carsten Maple\inst{1,2} \thanks{First three authors have equal contributions.}
\and
Graham Cormode\inst{1,2}
\and
Jiancong He\inst{3}}
\authorrunning{H. Kasyap et al.}
\institute{The Alan Turing Institute London, UK \\
\email{\{hkasyap,uatmaca\}@turing.ac.uk} \and
University of Warwick, UK \\ \email{\{cm,g.cormode\}@warwick.ac.uk}
\and
The Hongkong and Shanghai Banking Corporation Limited (HSBC), China\\
\email{winston.j.c.he@uni-heidelberg.de}}
%
%\end{comment}
\maketitle              % typeset the header of the contribution
\begin{abstract}
Financial institutions rely on data for many operations, including a need to drive efficiency, enhance services and prevent financial crime. Data sharing across an organisation or between institutions can facilitate rapid, evidence-based decision-making, including identifying money laundering and fraud. However, modern data privacy regulations impose restrictions on data sharing. For this reason, privacy-enhancing technologies are being increasingly employed to allow organisations to derive shared intelligence while ensuring regulatory compliance. 

This paper examines the case in which regulatory restrictions mean a party cannot share data on accounts of interest with another (internal or external) party to determine individuals that hold accounts in both datasets. The names of account holders may be recorded differently in each dataset. We introduce a novel privacy-preserving scheme for fuzzy name matching across institutions, employing fully homomorphic encryption over MinHash signatures. The efficiency of the proposed scheme is enhanced using a clustering mechanism. Our scheme ensures privacy by only revealing the possibility of a potential match to the querying party. The practicality and effectiveness are evaluated using different datasets, and compared against state-of-the-art schemes. It takes around 100 and 1000 seconds to search 1000 names from 10k and 100k names, respectively, meeting the requirements of financial institutions. Furthermore, it exhibits significant performance improvement in reducing communication overhead by 30–300 times.

\end{abstract}

\section{Introduction}
%1st Parag: Why data sharing is required for fraud detection
%2nd Parag: Use Case
%3rd Parag: SOTA
%4th Parag: Our paper
%5th Parag: Contributions

Financial crime is a significant threat to global safety and security. According to the Credit Industry Fraud Avoidance System (CIFAS), over 200,000 fraud cases were recorded in the UK’s National Fraud Database during the first half of 2024, with a 15\% increase compared to the same period in 2023 \cite{cifas2024fraudscape}. The United Nations Office on Drugs and Crime estimates that the annual amount of money laundered worldwide ranges from 2\% to 5\% of global GDP, which is 800 billion to 2 trillion US dollars \cite{unodc2024moneylaundering}. When a suspicious transaction or pattern is detected by a bank, the activity and associated accounts are typically reported to the bank's Financial Intelligence Unit (FIU) after an initial investigation \cite{FIU2024}. If the FIU determines that the activity is suspicious and potentially linked to a financial crime, it files a Suspicious Activity Report (SAR) \cite{network2003guidance}.

%Furthermore, detecting financial crimes is particularly challenging due to the complexity and cross-border distribution of the data across financial institutions, regulatory bodies, and law enforcement agencies. Thus, efficient sharing of financial intelligence is essential to enable rapid responses to combat these crimes.

%Efficient data sharing is increasingly vital for gaining the intelligence to prevent financial crime, considering the growing complexity of financial activities in the global economy. 

%Financial institutions typically report suspicious or unusual financial activities to Financial Intelligence Units (FIUs) for analysis and dissemination \cite{FIU2024}. Platforms like the Joint Money Laundering Intelligence Task Force in the UK \cite{NECC2024} and the Financial Crime Enforcement Network Exchange in the US \cite{FinCEN2024} have been implemented to facilitate collaboration between financial institutions, law enforcement agencies, and regulatory bodies. However, international banks would benefit from sharing information internally about potentially suspicious activities before revealing them to FIUs. This often happens during level 2 internal review, which enables swift and higher-confidence action to file a suspicious activity report (SAR). Advanced techniques, such as entity resolution \cite{kopcke2010evaluation}, enable banks to accurately identify and link records referring to the same real-world entities across different jurisdictions while ensuring compliance with privacy regulatory requirements.

The demand for timely fraud detection capabilities is increasing for banks, because the recent regulations mandate the authorized push payment fraud victims to be reimbursed within five days, with the cost shared equally between the sending and receiving payment service providers \cite{Sullivan2024}. Collaborative platforms such as the UK's Joint Money Laundering Intelligence Task Force (JMLIT) \cite{NECC2024} and the US Financial Crimes Enforcement Network Exchange (FinCEN Exchange) \cite{FinCEN2024} have been established to facilitate coordination between financial institutions, law enforcement agencies, and regulatory bodies. However, to meet the increasing regulatory demands, banks must enhance their internal financial intelligence capabilities, with a focus on early detection of suspicious activities. Advanced techniques, such as entity resolution \cite{kopcke2010evaluation}, enable banks to accurately identify and link records referring to the same real-world entities across different jurisdictions while ensuring compliance with privacy regulatory requirements.

%However, before sharing the potentially suspicious activity with FIUs, international banks (particularly branches of the same bank across demographics) will benefit from knowledge of potential threats for taking swift action while following regulatory compliance. Such knowledge can be obtained through entity resolution techniques \cite{kopcke2010evaluation}. %It is also referred to as record linkage or object matching and is designed for identifying records referring to the same real-world entity \cite{kopcke2010evaluation}.

The problem of entity resolution is a challenging one, due to the lack of distinct identifiers and the potential for transcription errors. 
%Entity resolution is an important task in data management~\cite{yakout2009efficient}, with applications in various sectors, including finance~\cite{carneiro2017data}. 
%and healthcare~\cite{hassan2017multimedia}. %It is used to improve the accuracy, efficiency, and security of transactions, while in healthcare, 
%It is employed for data cleaning and integration. It is also used to improve the accuracy, efficiency, and security of transactions in finance.
%Entity resolution is a critical task in data management. It has seen numerous applications in sectors like finance~\cite{carneiro2017data} and healthcare~\cite{hassan2017multimedia} to enhance the accuracy, efficiency, and security of transactions~\cite{yakout2009efficient}. 
%Particularly in scenarios lacking distinct identifiers, these systems are vital for mitigating 
%In financial services, it is common to have the risks associated with misidentification, fraud, and potential security breaches. 
In financial services, it is common to encounter variations of customer names in different forms, including misspellings, abbreviations, inconsistent formatting of first, middle, and last names, nicknames, variations in case usage, acronyms, leading and trailing spaces, and other similar variations~\cite{ansolabehere2017adgn}. This can create multiple identities with different names for the same customer, which may result in extra time and research spent identifying customer accounts and corresponding risks. Furthermore, a customer could maliciously use name variations to avoid detection by fraud prevention or anti-money laundering systems~\cite{gupta2020identity}. Private set intersection (PSI) and fuzzy (i.e., approximate) PSI methods enable joint computation for finding exact and similar items, respectively~\cite{chakraborti2023distance,essex2019secure,khurram2020sfour,uzun2021fuzzy,weicryptographically}.

Fuzzy PSI entails a private computation of the distance between two items. This distance is then compared to a predetermined similarity threshold to determine a match. Privacy Enhancing Technologies (PETs), such as Secure Multi-Party Computation (SMPC) or Homomorphic Encryption (HE), are commonly employed for private similarity computation~\cite{essex2019secure,weicryptographically}. SMPC-based approaches have been proposed for fuzzy name matching in recent studies~\cite{khurram2020sfour,weicryptographically}. However, these solutions allow both parties to learn the matching items (or some metadata), and also incur huge communication costs \cite{chakraborti2023distance}.
%This does not meet the requirements of the scenario we consider, where one party seeks to query another party privately. 
%We also require that the responding party remains unaware of the query's content and any potential matches. 
Furthermore, existing solutions tend to prioritize recall, while precision can be of greater significance in the financial context. A high precision means that the matching results are mostly correct, indicating fewer false positives. This is crucial when a mismatch cost is high in financial contexts, as mistakenly linking two different individuals could lead to serious errors, indirectly resulting in significant reputational harm.
%An increased occurrence of false positives may not only lead to privacy breaches but 
%It also jeopardises compliance with regulatory standards, which indirectly results in significant reputational harm.

%Protocols such as circuit-based techniques, oblivious polynomial evaluation using homomorphic encryption, and bloom filters are utilised for PSI. 

%However, these approaches are not sufficient to address the regulatory compliance in sensitive applications such as finance, because they publish the matching results to both parties involved. Thus, there is a need to develop privacy-preserving fuzzy name matching, utilising emerging Privacy Enhancing Technologies (PETs).% for applications such as cross-border fuzzy name matching.

HE-based approaches can ensure the privacy of sensitive information throughout the entire operation. This makes HE particularly attractive for cross-border financial data-sharing use cases, subject to diverse data privacy regulations. % since it reduces the risk of data breaches. 
HE simplifies compliance with these regulations by eliminating the need to decrypt data, even during processing~\cite{bed2016privately,calapodescu2017compact,chakraborti2023distance,wang2018privacy,uzun2021fuzzy}. 
However, a vital aspect to consider is the increasing computation cost, which introduces challenges in deploying HE-based approaches. 
HE-based solutions presented thus far~\cite{essex2019secure} have primarily utilised additive homomorphic encryption such as Paillier HE, based on the Decisional Composite Residuosity problem~\cite{orlandi2021rise}. However, recent advancements in HE has shifted focus to fully homomorphic encryption (FHE) based on lattice-based cryptography, such as CKKS (Cheon-Kim-Kim-Song) HE, which offers enhanced quantum resistance and is more adept at handling complex operations involving real or complex numbers~\cite{cheon2017homomorphic}. %In the context of fuzzy name matching in finance, where precision and security are paramount, the CKKS scheme's ability to handle real or complex numbers and its robustness against quantum computing threats make it an appropriate choice for secure computations.

%\noindent\textit{Challenges.} The major concerns in developing a privacy-preserving fuzzy PSI solution, which we aim to address, are: (1) high computation; (2) high communication; (3) efficiency and scalability; (4) perfect privacy and (5) compliance.
%\noindent\textit{Proposal.} 

This paper introduces a novel privacy-preserving fuzzy name matching scheme that offers a practical solution with a formal privacy guarantee. %for this real-world problem in the financial sector. 
Our scheme utilizes the cosine similarity of the signatures produced by MinHash algorithm (Section~\ref{sec-back}), %employing an effective LSH technique for textual data 
within the context of CKKS HE. Cosine similarity is preferred over other similarity metrics because of its independence over the length of records. %, considering an honest-but-curious threat model. 
Further, we utilise clustering to address the computational and communication overheads associated with using HE. 
Clustering is implemented using cosine similarity based K-Means, in offline on the responder side. Clustering \textit{significantly} reduces the search time, as the encrypted search operation is performed column-wise. A column has one element from each cluster, which is reduced to a single name from the closest cluster before performing a similarity check. It can also be parallelized, as operations over columns are independent.
%All the operations are performed in the encrypted domain, thus leaking no information to the responder. %And, if the name lies in the initial columns, it might only take a few seconds. 
The querier only learns the possibility of a potential match, without even learning the closeness. Thus, the scheme achieves perfect privacy. We present the implementation of the proposed scheme to validate its practicality for processing up to 1 million records. It has also been scaled for \textit{m-to-n} unbalanced searching ($m\ll n$), using batching in HE, which results in only a partial increase (not linear) in computational and communication overheads.
Clustering reduces memory consumption due to the column-wise operations in the encrypted domain, and also reduces communication overhead by a huge margin. For example, HE-based similarity search incurs 314GB of communication overhead if 100 names are searched serially (without clustering) from a million names, since all the similarity results are returned in the encrypted domain. 
Clustering reduces the overhead by orders of magnitude to 1-2GB. While clustering brings a drop in recall by a small amount, it preserves perfect precision. The key contributions of this paper are as follows: 

%It takes 13885 seconds to securely search a fuzzy name from 1 million records serially, but reduces to 4625 and 9250 seconds to search from 500k and 800k names, respectively, using clustering. Similarly, it incurs 175 GB of communication overhead in linear secure search, which reduces to 555 MB in the worst case, using clustering. However, clustering affects the accuracy by a small margin.
%\smallskip

%The key contributions of this paper are outlined as follows: 

\begin{itemize}
    \item We propose a novel privacy-preserving fuzzy name matching scheme by computing encrypted cosine similarities using CKKS encryption over MinHash signatures of names. It allows learning of a potential match only to the querying party, without revealing any other information to either party. % similarities for threshold fuzzy string matching.
    \item We enhance the practicality of the scheme by integrating clustering based on cosine similarity. It reduces the search space and facilitates quicker matching by performing column-wise operations. %It incurs a minor cost in recall while maintaining precision. 
    %While clustering brings a drop in recall by a small amount, it preserves perfect precision. 
    Further, it significantly reduces the computation and communication overheads. % by almost 30-300$\times$.  
    \item Evaluation of the scheme is presented in terms of accuracy, privacy and performance through implementation over different datasets and settings. 
    The empirical results, alongside our theoretical analysis, demonstrate the efficacy of the proposed solution. 
%    provide a shred of evidence for the proposed conjecture. The theoretical analysis supports the claim.
\end{itemize}

%The rest of this paper is organised as follows...
\section{Background and Related work}\label{sec-back}

\textbf{MinHash.} MinHash is particularly effective for estimating similarities between sets in high-dimensional spaces. It uses a random permutation function \(\pi\), which maps a universal set \(\Omega\) onto itself. Given a specific set \(u\), which is a subset of \(\Omega\), MinHash applies this permutation function \(\pi\) to each element within \(u\). The core idea of MinHash is to identify the smallest value resulting from these permutations. Formally, this is expressed as \(h_{\text{min}}^{\pi}(u) = \min\{\pi(x) \mid x \in u\}\), where \(h_{\text{min}}^{\pi}(u)\) denotes the MinHash value for the set \(u\)~\cite{broder1997syntactic}. This minimum value effectively captures a signature of the set for similarity comparisons, making MinHash a powerful tool for approximate nearest neighbor searches in large and complex datasets~\cite{wu2020review}. 

\noindent\textbf{Homomorphic Encryption.} Conventional encryption is vital for protecting data confidentiality while it is stored and in transit, but typically requires decryption to allow computations. %~\cite{mcmillan2013apple}. 
In 1978, Rivest introduced the concept of homomorphism for encryption, enabling computations on encrypted data without decryption, leading to numerous research efforts in homomorphic encryption (HE) schemes~\cite{rivest1978data}. %HE schemes come in different variants. Partial Homomorphic Encryption (PHE) allows specific operations like addition or multiplication; Somewhat Homomorphic Encryption (SHE) supports both but with operational limits; and Fully Homomorphic Encryption (FHE) schemes enable unrestricted polynomial functions on encrypted data~\cite{gentry2009fully,pulido2021privacy}. 
Existing schemes, namely BGV~\cite{brakerski2012fully} and BFV~\cite{brakerski2011fully}, perform arithmetic on integers, while the CKKS scheme~\cite{cheon2017homomorphic} enables approximate arithmetic on real or complex numbers. %All these schemes are rooted in the ring-learning-with-errors (RLWE) assumption, ensuring their security. Despite variations in their encryption techniques, they share a common data structure (such as polynomials) for representing ciphertexts. They all feature analogous fundamental homomorphic operations and employ comparable methods for maintaining ciphertext integrity, such as key switching. For this study, we employ CKKS due to its ability to perform approximate homomorphic computations across both the real and complex number domains.
The security of the CKKS scheme relies on the Ring Learning With Errors (RLWE) problem. It incorporates Gaussian noise primarily in encryption and key switching.
It addresses managing precision and approximation errors, particularly in operations like rescaling and encoding, which are vital to deal with to maintain the error within acceptable limits for the application. We refer to Appendix~\ref{sec:he} for more details.

\subsection{Related Work}
% fix any redundancies in related work and avoid strong statements

%This section discusses the state-of-the-art works in the fuzzy private set intersections. We refer to Appendix~\ref{sec:prelim} for preliminaries.

Privacy-preserving record linkage (PRL) represents a significant and common challenge in various application fields, such as finance and healthcare~\cite{gkoulalas2021modern,vatsalan2017privacy}. The earlier techniques primarily focused on identifying records with the same identifiers across different datasets, known as PSI~\cite{freedman2005keyword}. A range of protocols have been developed for PSI, including circuit-based protocols~\cite{pinkas2018efficient,chandran2021efficient}, oblivious polynomial evaluation enabled by homomorphic encryption for enhanced security and privacy~\cite{calapodescu2017compact,chen2018labeled,hu2023fully}, key agreement~\cite{jarecki2010fast}, and Bloom filters~\cite{debnath2015secure}. However, traditional PSI protocols typically lack the functionality to identify similar but not identical records. Such similarities often arise from data capture errors, particularly in genomics, surveillance, and finance. %These challenges highlight the importance and relevance of the topic.

Fuzzy PSI protocols are formed by employing a `closeness function' to assess the degree of similarity between records. 
Cryptographic techniques like HE~\cite{bed2016privately} or SMPC~\cite{indyk2006polylogarithmic} are integrated to ensure privacy. A predetermined distance threshold is commonly incorporated to assess the degree of similarity between the records~\cite{chakraborti2023distance}. Uzun et al.~\cite{uzun2021fuzzy} developed a protocol for computing private intersections of biometric data with sublinear communication scalability as dataset sizes increase, mainly for real-time surveillance scenarios. However, these techniques also consider identifiers (labels) associated with data. Thus, they fall in the category of fuzzy labeled PSI (FLPSI). They also have a high communication complexity of $\mathcal{O}(n^{2})$, given $n$ elements, held with both parties. Further, the subprotocol in~\cite{chakraborti2023distance} reveals both parties’ inputs when matched, revealing the result in pairs rather than only to the querier. These are based on Hamming distance-aware FLPSI, while our use case is suitable for cosine similarity distance metric.

%In the finance sector, regulatory compliance dictates specific requirements; notably, the responding party must remain unaware of matching items/scores. HE-based protocols offer significant privacy benefits in PRL applications. Neither party should gain any knowledge about the recorded details of any items held by the other party. This ensures a stringent level of privacy for sensitive financial data. Thus, the focus is shifting from standard PRL to secure, private searches. 

%A major limitation of SMPC-based PRLs~\cite{khurram2020sfour,weicryptographically,chakraborti2023distance} is that they make the intersection of data accessible to both parties. This can unintentionally expose the querier's search criteria and underlying intentions by analysing the query's structure, such as the query range~\cite{ling2023p}. However, HE-based protocols can keep the responder unaware of any matching results, thereby preserving the complete privacy of the querier's inquiry, which is suitable for compliance requirements. 
%Applying HE to string data requires a multi-step process, primarily due to HE's inherent design for numerical operations. The first step is converting the string into a numerical format~\cite{hahn2018practical}, such as binary encoding, or assigning a number to each character based on encoding schemes like ASCII or Unicode. These privacy benefits provide reassurance and confidence in the effectiveness of HE-based protocols.

Fuzzy PSI protocols typically involve two primary phases: blocking and matching~\cite{karakasidis2011secure,weicryptographically}, where the HE-based evaluation is employed for the matching phase. For blocking, LSH approaches are most often used for their ability to incorporate a numeric representation of a record~\cite{gyawali2020deduplication}. Unlike traditional hash functions, LSH aims to hash similar inputs into the same hash value, encouraging collisions such as grouping all similar names into one single digest value or into lists of digests that share common elements~\cite{adir2022privacy,khurram2020sfour}. However, there is a potential risk if LSH hashes expose private information about the input data, as highlighted by Turati et al.~\cite{turati2023locality}. Thus, LSH hashes are commonly processed using a privacy-preserving mechanism in the matching phase of PSI protocols~\cite{weicryptographically}. Our approach not only ensures compliance with stringent privacy regulations in finance but also %addresses the limitations of existing PRL methods for regularity compliance in finance, 
balances the need for high precision and recall, thereby reducing false positives and enhancing data utility.

%Table \ref{tab:relatedwork} to be completed...

\begin{comment}

\begin{table*}[!htbp]
\centering
\caption{Related Work}
\label{tab:relatedwork}
\begin{tabular}{|c|c|c|c|c|c|c|c|}
\hline
Study & Problem Addressed & Method & Input Data & Exact or Fuzzy & Metrics & Scalability & Limitation \\
\hline
\cite{uzun2021fuzzy} & Private Matching & \parbox[t]{2cm}{FHE \\ Garbled Circuits \\ Secret Sharing} & \parbox[t]{2cm}{Biometric} & Fuzzy & Euclidean Distance & High & \\
\hline
\cite{chakraborti2023distance}& Private Linkage &\parbox[t]{2cm}{AHE\\} &\parbox[t]{2cm}{Biometric\\ Network logs} & Fuzzy & \parbox[t]{2.5cm}{Minkowski distance \\ Hamming distance} & High & \\
\hline
\cite{essex2019secure}& &\parbox[t]{2cm}{} & & & & & \\
\hline
\cite{adir2022privacy}& &\parbox[t]{2cm}{} & & & & & \\
\hline
\cite{weicryptographically}& &\parbox[t]{2cm}{} & & & & & \\
\hline
\cite{khurram2020sfour}& &\parbox[t]{2cm}{} & & & & & \\
\hline
\end{tabular}
\end{table*}
\end{comment}

\section{Problem Definition}

\textbf{Problem Statement. } Consider two distinct private datasets managed by separate organisations \textit{A} and \textit{B}. Each record in both datasets represents an individual (or object) identified by their name and additional information such as date of birth. %Dataset in $A$ contains $m$ records, while dataset in $B$ contains $n$ records, and each record can be denoted as $r_i = {n_i, d_i}$. 
While these datasets may contain information about the same individuals, they might contain errors due to misrecording or intentional manipulation, resulting in inaccuracies.
%The objective of this study is to securely search for a record ($r^A_i$) from \textit{A} through all the records of \textit{B}, retrieving all those that correspond to the same real-world identity. 
The objective of this study is to find a potential match for a record ($r^A_i$) from \textit{A} in all the records of \textit{B}. %, retrieving all those that correspond to the same real-world identity. 
The potential match is decided using a similarity function.
%The matching function relies on a distance function due to the presence of inaccuracies within the datasets. 
Given records $r^A_i$ and $r^B_i$, let a similarity function be $\operatorname{Sim}(r^A_i,r^B_i): \rightarrow [0, 1]$ and a threshold value $\tau > 0$. Then the matching function returns a match if $\operatorname{Sim} \geq \tau$. %, and the pair of $(r^A_i,r^B_i)$ is declared to correspond to matching records. 
It should ensure that no additional information is disclosed to either organisation, apart from the possibility of a potential match. 
%\smallskip

\begin{comment}
The proposed system processes the names and associated identifiers to find if a similar (exact or fuzzy) record exists in another database. It should ensure the security of data by encrypting the flow of information and implementing strict access controls. The primary objective of the system is to efficiently match the records, all while maintaining the security and privacy of the query and the records held by another party. In the context of information security, the CIA triad—confidentiality, availability, and integrity—are the vital components of the system. Specifically:
\begin{itemize}
    \item Confidentiality: To ensure sensitive information is accessible only to those authorised to access it by implementing adequate access controls and authentication mechanisms to prevent unauthorised access to data and using encryption both in transit and at rest to protect sensitive data, such as personal identifiers used in the matching process.
    \item Availability: To ensure information and resources are accessible to authorised users when needed by implementing redundant data storage solutions to prevent data loss and regular security updates to maintain the system's operational functionality. \textbf{I don't think we are achieving this.}
    \item Integrity: To ensure data verification throughout its lifecycle through mechanisms including digital signatures, version control, and audit trails. \textbf{I don't think we are achieving this.}
\end{itemize}
\end{comment}
\noindent\textbf{Threat Model.} This study assumes that the system's security is adequately safeguarded against external threats. 
The proposed approach primarily addresses internal privacy concerns between $A$ and $B$. 
Thus, we consider a semi-honest (i.e., honest-but-curious) threat model where participating entities comply with the steps of the scheme and do not corrupt their inputs, yet they may attempt to gather/infer as much information as possible without deviating from or altering any aspect of the scheme~\cite{lindell2005secure,ranbaduge2015clustering}. %In a cross-border data sharing scenario, the proposed scheme is implemented between two branches of the same entity, operating under a mutual trust. 
Both branches execute the protocol using identical pre-processing steps and parameters for their respective local data repositories. 
%These entities cannot share the data due to regulatory mandates. 
The scheme enables that only the querying party gains insights from the similarity evaluation, while ensuring that the responding party does not acquire any information regarding the queries or the respective evaluation results. The querying party should also not be able to infer anything more than learning a potential match.

\section{Proposed Scheme}
%\noindent\textcolor{blue}{ CM: Whenever there is an ``identical case" we never miss!!! --- We never miss.} \\

%\noindent\textcolor{red}{Cormode: Can we use Additive HE or other FHE methods? Maybe we can make float integer by scaling factors, can we? --- Quantum resistance of CKKS added to the introduction.}\\
%%%%%%%%%%%%%%%%%%%%%%%%%%

This section details the proposed scheme for privacy-preserving fuzzy name matching. We begin by describing the system architecture and providing a clear understanding of its foundation. Then, we detail the steps, encompassing local data preprocessing, the derivation of MinHash signatures, private similarity measurement, and adjustments for the system's viable real-world implementation.

\begin{figure*}[t!]
  \centering
  \includegraphics[width=1\textwidth]{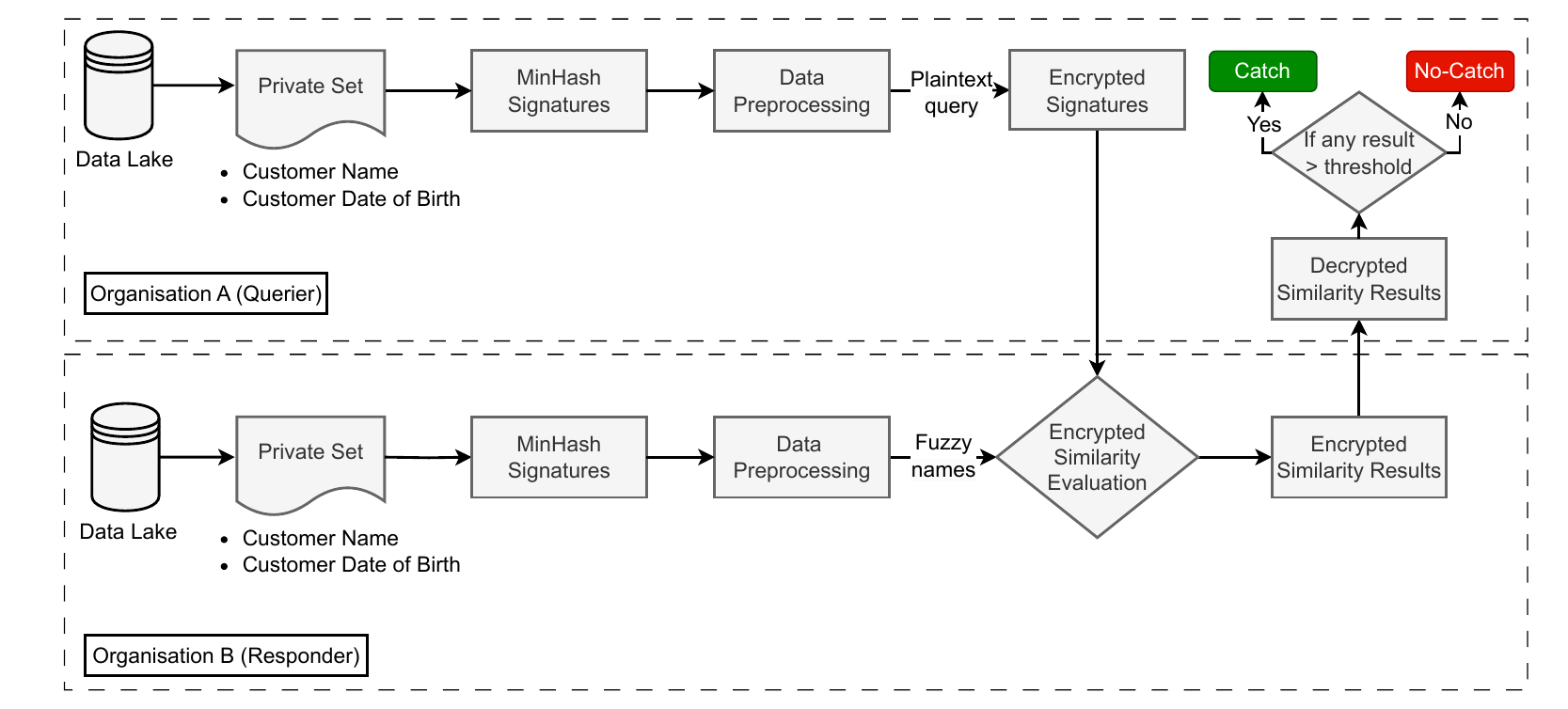}
  \caption{Workflow of the proposed privacy-preserving fuzzy name matching.% in anti-money laundering systems; the homomorphically encrypted MinHash signatures from the querier are shared with the responder to conduct encrypted similarity comparisons. %, and the encrypted results are returned. 
  %The querier decrypts the encrypted results to find `Catch' or `No-Catch', based on a threshold.
  }
  \label{fig:system_arch}
\end{figure*}

\noindent\textbf{Overview.} Fig.~\ref{fig:system_arch} demonstrates our proposed system architecture with two organisations, \textit{A} and \textit{B}. Both of them have their own private data lake. In this way, they develop their own private list of suspect individuals, which includes their name and additional information like date of birth. 
Since MinHash preserves the pattern of names, \textit{A} and \textit{B} encode names using MinHash and generate signatures. \textit{A} is now looking for a possible match in organisation \textit{B}. \textit{A} normalises the generated MinHash signatures (to query) and then employs CKKS encryption to securely share it with \textit{B}. \textit{B} computes the dot product with the received encrypted query from \textit{A} with all the normalised MinHash signatures held by \textit{B}, serially. It should be noted that the dot product (sum of component-wise products) is a supported operation in fully homomorphic encryption. 
Here, the dot product between two normalised vectors represents the cosine similarity between them. Finally, \textit{A} receives back the encrypted similarity result and decrypts it to determine whether the name is present in \textit{B}'s data lake. Now, we discuss each component in detail.

\begin{comment}
Algorithm~\ref{alg:lsh} outlines the LSH signature generation methodology. The following example is designed to demonstrate the steps involved in the generation of LSH signatures in two distinct entities, \textit{A} and \textit{B}. The dataset of \textit{A} contains a list of names as $N_A = \{\text{Mary Janes}, \text{John Doe}, \text{Alice White}\}$, while \textit{B}'s dataset contains $N_B = \{\text{Emma Johnson}, \text{David Brown}, \text{Marie Jones}\}$. The datasets share a commonality in the names ``Mary Janes” in $N_A$ and ``Marie Jones” in $N_B$, emphasising the need for careful signature generation. Both \textit{A} and \textit{B} utilise the same LSH technique, using the same parameters to effectively create similar signatures for the records in their respective datasets.
\end{comment}

%The dataset of A is defined as $D_A = \{\text{Mary Janes}, \text{John Doe}, \text{Alice White}\}$, while B's dataset is $D_B = \{\text{Emma Johnson}, \text{David Brown}, \text{Marie Jones}\}$. The datasets share a commonality in the names ``Mary Janes” in $D_A$ and ``Marie Jones” in $D_B$, emphasising the need for careful signature generation. Both A and B utilise the same LSH technique with the same parameters to effectively create signatures for the records in their respective datasets.

%Algorithm~\ref{alg:lsh} outlines the MinHash signature generation methodology. First, 

\noindent\textbf{Dataset Encoding.} The sets of shingles (i.e., $n$-grams $\{g_1, g_2, ..., g_n\}$) are generated for the records in \textit{A} and \textit{B} which are two distinct entities. These sets contain all possible substrings of $n$ consecutive  characters, including spaces. The hash value $h_i=H(g_i)$ for each $g_i$ is calculated using a hash function $H$. Then, MinHash technique is applied to find the minimum hash value among these $n$-grams, which represents the MinHash signatures as $\operatorname{min}(H(g_i))$. The complete procedure can be referred to in Algorithm~\ref{alg:lsh} of Appendix~\ref{sec:algorithms}.
\begin{comment}
For example, using $3$-grams, the breakdown for each name in the datasets is as follows:
\begin{itemize}
    \item $D_A$: \{[`Mar', `ary', `ry ', `y J', ` Ja', `Jan', `ane', `nes'], [`Joh', `ohn', `hn ', `n D', ` Do', `Doe'], [`Ali', `lic', `ice', `ce ', `e W', ` Wh', `Whi', `hit', `ite']\} 
    \item $D_B$: \{[`Emm', `mma', `ma ', `a J', ` Jo', `Joh', `ohn', `hns', `nso', `son'], [`Dav', `avi', `vid', `id ', `d B', ` Br', `Bro', `row', `own'], [`Mar', `ari', `rie', `ie ', `e J', ` Jo', `Jon', `one', `nes'] \}
\end{itemize}
\end{comment}

The size of the $n$-gram is critical in determining the granularity of similarity detection. Here, the generated $3$-grams are subject to hashing using an approximate function, such as SHA-256, to produce a unique hash value. Subsequently, a set of random permutation functions is created. Increasing the number of permutation functions enhances the precision of the resultant MinHash signatures. Simultaneously, this increases the size of the signatures, thereby enhancing the accuracy, but also increasing computational cost and storage requirements. Following this, a list of buckets is constructed, which serves to constrain the dimensions of the hash values. This constraint plays a pivotal role in determining the probability of hash collisions. 
Adjusting the maximum hash value parameter expands the range of possible hash values, lowering the probability of collisions occurring. 
It can be fine-tuned to balance between accuracy and computational efficiency in the hashing process. However, performing the search in an encrypted domain serially %, as demonstrated by Fig.~\ref{fig:system_arch},  
 incurs high costs. 
Typically, \textit{B} may have a huge list of names, say, $|N_B| \approx 10^6$. 
Computing and returning encrypted similarity scores for $10^6$ names is a computation and communication-intensive task. 
Moreover, it will reveal the similarity scores associated with each name owned by \textit{B}.% to \textit{A}.

\begin{algorithm}[!t]
\caption{Private Fuzzy Name Matching}
\label{alg:scheme1}
%\scalebox{0.95}{
\begin{minipage}{1\textwidth}
\KwData{Query Signature $\vec{N_A}$, %\\\setlength\parindent{27pt} 
Cosine Similarity Threshold $\tau$}
\KwResult{catch (1 if match else 0)}
\textbf{Function} CompareToCentroids $(\vec{C_k}$, $E(\hat{N_{As}}))$\;
\Begin{
    $E(\text{sim\_scores}) \gets \operatorname{DotProduct}(\vec{C_k}$, $E(\hat{N_{As}}))$\;
    Serialise and send the encrypted similarity scores $E(\text{sim\_scores})$ of the encrypted query (normalised and scaled MinHash signature) with the centroids;}
\textbf{Function} ColumnWiseMatching $(C$, $E(\hat{N_{A}})$, $E(\vec{S}))$\;
\Begin{
    \For{each column in $C$}{
        $E(\hat{N}_{B_{ij}}) \gets \operatorname{DotProduct}(E(\vec{S}), \hat{N}_{B_{column}})$; \Comment{This results in the encrypted name from the possible matching cluster ($i$) in this column ($j$)}\;
        $E(\text{cos\_score}) \gets \operatorname{DotProduct}(E(\hat{N}_{B_{ij}}), E(\hat{N_{A}}))$; \Comment{This is an in-place dot product resulting in the cosine similarity between the encrypted name and the encrypted query}\;

        $E(\text{score}) \gets \operatorname{CKKSAdd_{const}}(E(\text{cos\_score}), - \tau)$; \Comment{Subtract $\tau$}\;
        $E(\text{score}) \gets \operatorname{CKKSMult_{const}}(E(\text{score}), r)$; \Comment{Multiply with random $r \in \mathbb{Z}_p^*$}\;
        
        %$E(\text{score}) \gets r * (E(\text{cos\_score}) - \tau)$; \Comment{Subtract the threshold $\tau$ and multiply with a random number $r \in \mathbb{Z}_p^*$}\;
        Serialise and send $E(\text{score})$ of the encrypted name from the matching cluster in this column and the encrypted query;
    }
    }    
\Begin{
    $B$ performs clustering in offline to get clustered matrix ($C$) and centroids ($\vec{C_k}$)\;
    $A$ executes the below steps online\;
    $\hat{N_{A}} \gets \frac{\vec{N_A}}{|\vec{N_A}|}$; \Comment{Vector Normalisation}\;
    $\hat{N_{As}} \gets \operatorname{StandardScaler}(\hat{N_{A}})$; \Comment{Standardize by removing the mean and scaling to unit variance}\; 
    $E(\hat{N_{A}}) \gets $ Encrypt 
    $\hat{N_{A}}$ with the public key of \textit{A}\;
    $E(\hat{N_{As}}) \gets $ Encrypt 
    $\hat{N_{As}}$ with the public key of \textit{A}\;
    $E(\text{sim\_scores}) \gets $ Sends $E(\hat{N_{As}})$ to \textit{B} to invoke the function CompareToCentroids and receive encrypted similarities scores with the centroids\;
    $\text{sim\_scores} \gets $ Decrypts $E(\text{sim\_scores})$\;
    $\vec{S} \gets $ Prepares an indicator/sign (one-hot) vector with 1 for the most matching cluster and 0's for the remaining\;
    $E(\vec{S}) \gets $ Encrypts $\vec{S}$ with the public key of \textit{A}\;
    $E(\text{cos\_score}) \gets $ Sends $E(\hat{N_{A}})$ and $E(\vec{S})$ to \textit{B} to invoke the function ColumnWiseMatching and keeps receiving encrypted cosine similarities for each column\;
    $\text{score} \gets $ Decrypts $E(\text{score})$\;
    catch $\gets 1 $ \textbf{if any} $\text{score} > 0$
}
\end{minipage}
%}
\end{algorithm}

\subsection{Clustering-based Matching}
\label{V.B}
We integrate a clustering-based approach to enhance the performance of the scheme, in terms of computation, communication, memory, and time. The responding organisation ($B$) clusters its private dataset to reduce the search space for the query. The proposed clustering-based approach entails sending approximately $\sqrt{|N_B|}$ similarity scores. 
It reduces both the number of comparisons and communication overhead. Algorithm~\ref{alg:scheme1} outlines this scheme. Assume that \textit{A} has an encoded name, $\vec{N_A}$, with which it intends to perform fuzzy name matching with \textit{B}'s list of encoded names ($\vec{N_B}$). The following steps are executed.

%As, \textit{B} may have a huge list of names, it will require a huge comparison, which is not practical in terms of time, memory, computation and communication.

%The dataset in Organisation B has been partitioned into about $\sqrt{D}$ clusters, including a total of $D$ records. Clustering is a one-time procedure, and every new record added to Organisation B's dataset is assigned to the nearest cluster centroid, with the assignment based on the cosine similarity measure.

%\begin{enumerate}
    %\noindent
    \textbf{1.} Organisation \textit{B} forms approximately $\sqrt{|N_B|}$ (say $k$) clusters of MinHash signatures, ensuring that similar encoded values are grouped together. $B$ uses the K-means clustering algorithm, based on the cosine similarity measure. Clustering can be performed offline at $B$ before the search operation, as described in Algorithm~\ref{alg:clustering} of Appendix~\ref{sec:algorithms}. As a preprocessing step, $B$ performs normalization and standardization of the MinHash signatures before clustering. 
    Conceptually, we can imagine that $B$ creates a $k \times k$ matrix, where each row represents a distinct cluster. 
    However, it is impractical to cluster such that \textit{B} gets a perfectly square matrix. 
    For realistic datasets and distributions, 
    it is common that some clusters will collect a much higher number of names. 
    Thus, $B$ needs to pad all the clusters up to the maximum number of names in any cluster. This is all handled by the $\operatorname{Clustering}$ function called by Algorithm~\ref{alg:scheme1}, executed by \textit{B}.

    %\noindent
    \textbf{2.} Organisation \textit{A} normalises the encoded query vector ($\vec{N_A}$) to get $\hat{N}_A$. It also performs standardization to get $\hat{N}_{As}$. Then, \textit{A} encrypts $\vec{N_A}$ and $\vec{N_{As}}$ using its public key. First, \textit{A} sends $E(\hat{N}_{As})$ to \textit{B}, to match with centroids. This is performed by executing $\operatorname{CompareToCentroids}$ function of Algorithm~\ref{alg:scheme1}.

    %\noindent
    \textbf{3.} Organisation \textit{B} calculates the dot (inner) product of $E(\hat{N}_{As})$ with all the centroids ($\vec{C_k}$). This is an inner product between ciphertext and plaintext, as defined in Algorithm~\ref{alg:dotp} of Appendix~\ref{sec:algorithms}. Thus, B calculates a vector of $k$ similarity scores, encrypted with \textit{A}'s public key, as $E(\text{sim\_scores})$, and sends to \textit{A}.
    
    %\noindent
    \textbf{4.} Upon receiving the encrypted similarity scores with centroids, \textit{A} decrypts to get $\text{sim\_scores}$. Then \textit{A} prepares an indicator (sign) vector, which assigns a score of 1 to the highest matching centroid, while all other scores are set to 0. 
    This results in a one-hot vector ($\vec{S}$) of 0's and 1 of size $1 \times k$.

    %\noindent
    \textbf{5.} Organisation \textit{A} encrypts $\vec{S}$ with its public key and sends the encrypted indicator vector $E(\vec{S})$ along with the encrypted normalised query $E(\hat{N_{A}})$ to \textit{B}.

    %\noindent
    \textbf{6.} Organisation \textit{B} performs column-wise operations over the matrix prepared in step 1. Each column contains one element from each cluster. \textit{B} performs multiplication and addition operations to find dot/inner product between $E(\vec{S})$ and a column. This is an inner product between ciphertext and plaintext, as defined in Algorithm~\ref{alg:dotp}. After this operation, in the encrypted domain, \textit{B} gets the normalised MinHash encoded signature from the matching cluster (represented by 1 in the encrypted indicator vector). Since the result is encrypted by the public key of \textit{A}, \textit{B} does not learn about the position (cluster/row) of the name. For instance, 
\[ {E([0,0,1,0,0])~\times~}\begin{bmatrix}
    \hat{N}_{B_{11}} \\
    \hat{N}_{B_{12}} \\
    \hat{N}_{B_{13}} \\
    \hat{N}_{B_{14}} \\
    \hat{N}_{B_{15}}
\end{bmatrix}
= E([\hat{N}_{B_{13}}])
\]
%        \[\begin{aligned}
%        & {[\hat{N}_{B_{11}},} \\
%        & \hat{N}_{B_{12}}, \\
%        {E([0,0,1,0,0]) \quad \times } & \hat{N}_{B_{13}}, & = E([\hat{N}_{B_{13}}]) \\
%        & \hat{N}_{B_{14}}, \\
%5        & \hat{N}_{B_{15}}]
%        \end{aligned},\] 
where $\hat{N}_{B_{13}}$ represents the third name from the first column.

    %B multiplies its matrix of $\sqrt{D} \times \sqrt{D}$ with $E_{pk_{A}}(M)$, such that, the highest matching row (cluster) with potential similarity remains the same, and the other rows (clusters) become 0. However, we multiply $E_{pk_{A}}(M)$ with each column, for parallelization. A column represents one element from each cluster.

    %\item B sums column-wise elements in the matrix, which results in the names only from the potential matching cluster, then B finds cosine similarity of $E_{pk_{A}}(N)$ with each name.

    %\noindent
    \textbf{7.} \textit{B} finds the dot product of the output from the previous step to $E(\hat{N_{A}})$. 
    This is an inner product between ciphertext and ciphertext, as defined in Algorithm~\ref{alg:dotp}. This represents the encrypted cosine similarity $E(\text{cos\_score})$ between the querying name and the name from the matching cluster in this column.
    
    \textbf{8.} \textit{B} subtracts the threshold ($\tau$) from the $E(\text{cos\_score})$ and multiplies it with a random number ($r \in \mathbb{Z}_p^*$) to get $E(\text{score})$. Then, \textit{B} sends $E(\text{score})$ to \textit{A} for each column. Thus, the cosine similarities above the threshold remain positive, and others remain negative, after subtracting $\tau$. Also, since every score is multiplied by a different random number, no relation is preserved among the distribution of scores.
    Steps 6-8 can be referred to in the $\operatorname{ColumnWiseMatching}$ function of Algorithm~\ref{alg:scheme1}.

    %\noindent
    \textbf{9.} Organisation \textit{A} receives $E(\text{score})$ and decrypts it with its secret key. If the score is positive, then it can be considered as a potential approximate match present with \textit{B}.
%\end{enumerate}

In this way, \textit{A} and \textit{B} execute the above steps for privacy-preserving fuzzy name matching. 
After executing this protocol, \textit{A} only learns whether there exists a potential match or not.
%learns similarity to centroids, which can also be obscured by adding a random number. 
These similarity results do not reveal any closeness to the query. 
%This procedure leaks negligible information about the centroids, since many vectors can satisfy this similarity score. 
%Next, \textit{A} receives cosine similarities with the names from the matching cluster, which is the requirement. 
Meanwhile, \textit{B} does not learn anything throughout the whole process, neither about the query nor the response.

A novel feature of our scheme is that \textit{A} does not need to wait until it receives all scores from \textit{B}. 
There is a high probability that it may get a match (if it exists) from the results received only for a few columns. Hence, it can be understood that padding does not impact the performance of other (smaller) clusters. As we see empirically, it is common that 80-90\% of names are covered in only half of the columns. The remaining half only includes 5-10\% of names from one or two clusters with larger sizes. In summary, the proposed scheme provides secure, private, efficient, fuzzy name matching based on fully homomorphic encryption.
%In the subsequent section, we formalize the security properties and accuracy guarantees.
%also respecting regulatory compliance, to the best of our knowledge.
We refer to Appendix~\ref{sec:theoanal} for theoretical (privacy and correctness) analysis.

\section{Experimental Study}
This section evaluates the proposed privacy-preserving fuzzy name matching scheme over different datasets with multiple name encoding and clustering settings. The scheme has been empirically evaluated both for accuracy and performance measures. %Further, the evaluation results are analyzed to justify the selection of different parameters for fast and accurate fuzzy matching. 
The implementation of the proposed scheme uses a Python library TenSEAL\footnote{https://github.com/OpenMined/TenSEAL}, supporting homomorphic operations over tensors, including addition, and multiplication. These are supported as in-place operations, saving memory and time. It also supports batching, which enables packing multiple plaintexts in one ciphertext, enabling parallelization and significantly improving the performance. 
TenSEAL also has an implementation for serialization and deserialization, which helps evaluate the real-time communication cost. 
%The experiments have been benchmarked on Apple MacBook M2 Pro with 32 GB RAM.
%\smallskip

\noindent\textbf{Parameters.} It includes setting the parameters for data encoding, clustering and CKKS HE in TenSEAL. For data encoding, we configure (1) shingle\_size: the granularity of similarity depends on this; (2) num\_permutations: length of encoding, and a large number produces a more accurate encoding; however, it also increases the computational and communication overheads associated; (3) max\_hash: the size of the hash values is bound by this, which also affects the likelihood of hash collisions. For evaluation, (1) shingle\_size is set to 3, based on the existing studies; (2) num\_permutations as 50, 100 and 200; (3) max\_hash is set to 20-bit hash values, which is a large space to avoid collisions.

For clustering, we set the number of clusters close to the square root of the number of samples held by the responder. However, we investigate the impact with varying numbers of clusters. The number of iterations is set to 20. Before clustering, we do preprocessing: Normalization ($\hat{N} \gets \frac{\vec{N}}{||\vec{N}||}$) followed by Standardization using StandardScaler\footnote{https://scikit-learn.org/stable/modules/preprocessing.html} package.

For CKKS, we configure (1) scaling factor: which defines the precision of encoding for the binary representation; (2) polynomial modulus degree: a larger degree has higher security but increased computation and communication; (3) coefficient modulus sizes: a list of primes indicating the level of multiplication supported. For evaluation, (1) scaling factor is set as $2^{40}$; (2) polynomial modulus degree as 8192; and (3) coefficient modulus sizes as [60, 40, 40, 60].
%\smallskip

%homopolynomial modulus degree, which is a balancing factor for the security level and computational efficiency of the encryption. The precision of the computations, along with the range and accuracy of encrypted calculations, are defined by the bit sizes of the coefficients. These bit sizes also influence the size of the modulus chain. Additionally, the global scale is a factor in noise growth management during arithmetic operations on encrypted data. The following parameters are selected for CKKS in TENSEAL:\\

%\indent \text{poly\_modulus\_degree} = $8192$,\\
%\indent \text{coeff\_mod\_bit\_sizes} = $[60, 40, 40, 60]$,\\
%\indent \text{global\_scale} = $2^{40}$.\\

\noindent\textbf{Datasets.}
The experiment is conducted using three datasets, North Carolina Voter Registration (NCVR)~\cite{NCSBE}, library catalogue datasets~\cite{bodo2020shadow,simakis_2020} and US Census~\cite{us_census}. The details of datasets can be referred to in Appendix~\ref{sec:datasets}.

%\smallskip

%\noindent\textbf{Baseline Method:} We evaluate our method as stated in Fig.~\ref{fig:system_arch}, without clustering, and consider it as a baseline. The accuracy of the approach is not impacted due to encryption; thus, results are not reported without encryption.
%\smallskip

\noindent\textbf{Compared Approaches.} We consider state-of-the-art approaches for fuzzy private record linkage, HE-based~\cite{essex2019secure}, and SMPC-based~\cite{khurram2020sfour,weicryptographically} for comparison. However, our requirement is not a private fuzzy linkage but an unbalanced private fuzzy search. Thus, we compare in approximately similar settings.
%\smallskip

\noindent\textbf{Evaluation metrics.}
%To evaluate the efficacy and practicality of our proposed scheme, a suite of metrics is utilised. 
We measure the computation and communication costs to evaluate the practicality of the scheme. In terms of search efficiency, accuracy measures the overall correctness of the classifier by calculating the ratio of correct predictions against the total number of predictions made. Complementing accuracy, recall focuses on the classifier's capability to accurately identify positive instances out of all actual positive cases. Precision emphasises the accuracy of positive predictions made by the classifier. The F1 Score synthesises these metrics using a harmonised mean that effectively balances precision and recall. The sensitivity analysis is also conducted to observe the efficiency in handling increasing fuzziness in the records measured by Levenshtein distances.

\subsection{Results and Discussion}
\begin{figure}
    \centering
    %\resizebox{0.45\textwidth}{0.4\textwidth}{%
    \begin{tikzpicture}[thick, scale=0.6, every node/.style={transform shape}]
    \begin{axis}[
        %label style={font=\huge},         ticklabel style = {font=\Large},
        legend columns=2, 
        legend style={at={(0.5,0.5)},anchor=north,draw={none}},
        xlabel= Cosine Similarity Threshold,
        ylabel= Percentage,
        ymin=0, ymax=1,
        xticklabels={0.5,0.55,0.6,0.65,0.7,0.75,0.8,0.85,0.9,0.95},
        ytick={0,.20,.40,.60,.80,1},
        grid=both,
        xtick=data,
        xtick distance=1,
        table/x expr=\coordindex,
        ]

    \addplot[color=black,mark=+] table [x=Clusters, y=Accuracy, col sep=comma] {results/nc10k_threshold.csv};
    \addlegendentry{Accuracy}
    \addplot[color=red,mark=x] table [x=Clusters, y=Precision, col sep=comma] {results/nc10k_threshold.csv};
    \addlegendentry{Precision}
    \addplot[color=cyan,mark=*] table [x=Clusters, y=Recall, col sep=comma] {results/nc10k_threshold.csv};
    \addlegendentry{Recall}
    \addplot[color=blue,mark=o] table [x=Clusters, y=F1-Score, col sep=comma] {results/nc10k_threshold.csv};
    \addlegendentry{F1-Score}
    
    \end{axis}
    \end{tikzpicture}%}
    \caption{Varying cosine similarity threshold.}% on NCVR dataset (Settings: 10K - Clusters 200).}
    \label{fig:nc10k_threshold}
\end{figure}
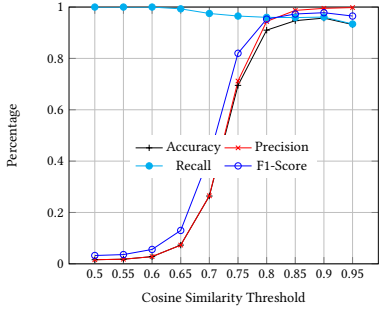

\noindent\textbf{Accuracy Evaluation.} We study the search efficiency of the proposed scheme and report the accuracy, precision, recall and F1-score under different settings. First, we perform an experiment to select a threshold for a cosine similarity score. Fig.~\ref{fig:nc10k_threshold} demonstrates the performance with varying cosine similarity from 0.5 to 0.95 on the NCVR dataset. It can be observed that recall is approximately 1 for threshold 0.65, but precision is near to 0. Recall and precision are nearly the same, from 0.8 to 0.9, while accuracy is also nearly the same at 0.9. However, recall drops when the threshold is 0.95. Thus, we chose 0.9 as the threshold for the remaining experiments.

\begin{multicols}{2}
\begin{figure}[H] 
\centering
\begin{subfigure}[b]{0.15\textwidth}
    \centering
    \resizebox{\textwidth}{\textwidth}{%
    \begin{tikzpicture}[thick, scale=1, every node/.style={transform shape}]
    \begin{axis}[
        label style={font=\huge},         ticklabel style = {font=\Large},
        legend columns=2, 
        legend style={at={(0.5,0.5)},anchor=north,draw={none},font=\Large},
        xlabel= Clusters,
        ylabel= Percentage,
        ymin=0, ymax=1,
        xticklabels={0,50,100,150,200},
        ytick={0, 0.20, 0.40, .60, .80, 1},
        grid=both,
        xtick=data,
        xtick distance=1,
        table/x expr=\coordindex,
        ]

    \addplot[color=black,mark=+] table [x=Clusters, y=Accuracy, col sep=comma] {results/nc10k_lsh50.csv};
    \addlegendentry{Accuracy}
    \addplot[color=red,mark=x] table [x=Clusters, y=Precision, col sep=comma] {results/nc10k_lsh50.csv};
    \addlegendentry{Precision}
    \addplot[color=cyan,mark=*] table [x=Clusters, y=Recall, col sep=comma] {results/nc10k_lsh50.csv};
    \addlegendentry{Recall}
    \addplot[color=blue,mark=o] table [x=Clusters, y=F1-Score, col sep=comma] {results/nc10k_lsh50.csv};
    \addlegendentry{F1-Score}
    
    \end{axis}
    \end{tikzpicture}}
    \caption{EL 50}
    \label{fig:nc10k_lsh50}
\end{subfigure}
\begin{subfigure}[b]{0.15\textwidth}
    \centering
    \resizebox{\textwidth}{\textwidth}{%
    \begin{tikzpicture}[thick, scale=1, every node/.style={transform shape}]
    \begin{axis}[
        label style={font=\huge},         ticklabel style = {font=\Large},
        legend columns=2, 
        legend style={at={(0.5,0.5)},anchor=north,draw={none}},
        xlabel= Clusters,
        ylabel= Percentage,
        ymin=0, ymax=1,
        xticklabels={0,50,100,150,200},
        ytick={0,.20,.40,.60,.80,1},
        grid=both,
        xtick=data,
        xtick distance=1,
        table/x expr=\coordindex,
        ]

    \addplot[color=black,mark=+] table [x=Clusters, y=Accuracy, col sep=comma] {results/nc10k_lsh100.csv};
    %\addlegendentry{Accuracy}
    \addplot[color=red,mark=x] table [x=Clusters, y=Precision, col sep=comma] {results/nc10k_lsh100.csv};
    %\addlegendentry{Precision}
    \addplot[color=cyan,mark=*] table [x=Clusters, y=Recall, col sep=comma] {results/nc10k_lsh100.csv};
    %\addlegendentry{Recall}
    \addplot[color=blue,mark=o] table [x=Clusters, y=F1-Score, col sep=comma] {results/nc10k_lsh100.csv};
    %\addlegendentry{F1-Score}
    
    \end{axis}
    \end{tikzpicture}}
    \caption{EL 100}
    \label{fig:nc10k_lsh100}
\end{subfigure}
\begin{subfigure}[b]{0.15\textwidth}
    \centering
    \resizebox{\textwidth}{\textwidth}{%
    \begin{tikzpicture}[thick, scale=1, every node/.style={transform shape}]
    \begin{axis}[
        label style={font=\huge},         ticklabel style = {font=\Large},
        legend columns=2, 
        legend style={at={(0.5,0.5)},anchor=north,draw={none}},
        xlabel= Clusters,
        ylabel= Percentage,
        ymin=0, ymax=1,
        xticklabels={0,50,100,150,200},
        ytick={0,.20,.40,.60,.80,1},
        grid=both,
        xtick=data,
        xtick distance=1,
        table/x expr=\coordindex,
        ]

    \addplot[color=black,mark=+] table [x=Clusters, y=Accuracy, col sep=comma] {results/nc10k_lsh200.csv};
    %\addlegendentry{Accuracy}
    \addplot[color=red,mark=x] table [x=Clusters, y=Precision, col sep=comma] {results/nc10k_lsh200.csv};
    %\addlegendentry{Precision}
    \addplot[color=cyan,mark=*] table [x=Clusters, y=Recall, col sep=comma] {results/nc10k_lsh200.csv};
    %\addlegendentry{Recall}
    \addplot[color=blue,mark=o] table [x=Clusters, y=F1-Score, col sep=comma] {results/nc10k_lsh200.csv};
    %\addlegendentry{F1-Score}
    
    \end{axis}
    \end{tikzpicture}}
    \caption{EL 200}
    \label{fig:nc10k_lsh200}

\end{subfigure}

\caption{Accuracy on NCVR dataset (10k-10k). EL (Encoding Length).}
\label{fig:nc}
\end{figure}

\begin{figure}[H] 
\centering
\begin{subfigure}[b]{0.15\textwidth}
    \centering
    \resizebox{\textwidth}{\textwidth}{%
    \begin{tikzpicture}[thick, scale=1, every node/.style={transform shape}]
    \begin{axis}[
        label style={font=\huge},         ticklabel style = {font=\Large},
        legend columns=2, 
        legend style={at={(0.5,0.5)},anchor=north,draw={none},font=\Large},
        xlabel= Clusters,
        ylabel= Percentage,
        ymin=0, ymax=1,
        xticklabels={0,50,100,150,200},
        ytick={0,.20,.40,.60,.80,1},
        grid=both,
        xtick=data,
        xtick distance=1,
        table/x expr=\coordindex,
        ]

    \addplot[color=black,mark=+] table [x=Clusters, y=Accuracy, col sep=comma] {results/book_lsh50.csv};
    \addlegendentry{Accuracy}
    \addplot[color=red,mark=x] table [x=Clusters, y=Precision, col sep=comma] {results/book_lsh50.csv};
    \addlegendentry{Precision}
    \addplot[color=cyan,mark=*] table [x=Clusters, y=Recall, col sep=comma] {results/book_lsh50.csv};
    \addlegendentry{Recall}
    \addplot[color=blue,mark=o] table [x=Clusters, y=F1-Score, col sep=comma] {results/book_lsh50.csv};
    \addlegendentry{F1-Score}
    
    \end{axis}
    \end{tikzpicture}}
    \caption{EL 50}
    \label{fig:book_lsh50}

\end{subfigure}
\begin{subfigure}[b]{0.15\textwidth}
    \centering
    \resizebox{\textwidth}{\textwidth}{%
    \begin{tikzpicture}[thick, scale=1, every node/.style={transform shape}]
    \begin{axis}[
        label style={font=\huge},         ticklabel style = {font=\Large},
        legend columns=2, 
        legend style={at={(0.5,0.5)},anchor=north,draw={none}},
        xlabel= Clusters,
        ylabel= Percentage,
        ymin=0, ymax=1,
        xticklabels={0,50,100,150,200},
        ytick={0,.20,.40,.60,.80,1},
        grid=both,
        xtick=data,
        xtick distance=1,
        table/x expr=\coordindex,
        ]

    \addplot[color=black,mark=+] table [x=Clusters, y=Accuracy, col sep=comma] {results/book_lsh100.csv};
    %\addlegendentry{Accuracy}
    \addplot[color=red,mark=x] table [x=Clusters, y=Precision, col sep=comma] {results/book_lsh100.csv};
    %\addlegendentry{Precision}
    \addplot[color=cyan,mark=*] table [x=Clusters, y=Recall, col sep=comma] {results/book_lsh100.csv};
    %\addlegendentry{Recall}
    \addplot[color=blue,mark=o] table [x=Clusters, y=F1-Score, col sep=comma] {results/book_lsh100.csv};
    %\addlegendentry{F1-Score}
    
    \end{axis}
    \end{tikzpicture}}
    \caption{EL 100}
    \label{fig:book_lsh100}

\end{subfigure}
\begin{subfigure}[b]{0.15\textwidth}
    \centering
    \resizebox{\textwidth}{\textwidth}{%
    \begin{tikzpicture}[thick, scale=1, every node/.style={transform shape}]
    \begin{axis}[
        label style={font=\huge},         ticklabel style = {font=\Large},
        legend columns=2, 
        legend style={at={(0.5,0.5)},anchor=north,draw={none}},
        xlabel= Clusters,
        ylabel= Percentage,
        ymin=0, ymax=1,
        xticklabels={0,50,100,150,200},
        ytick={0,.20,.40,.60,.80,1},
        grid=both,
        xtick=data,
        xtick distance=1,
        table/x expr=\coordindex,
        ]

    \addplot[color=black,mark=+] table [x=Clusters, y=Accuracy, col sep=comma] {results/book_lsh200.csv};
    %\addlegendentry{Accuracy}
    \addplot[color=red,mark=x] table [x=Clusters, y=Precision, col sep=comma] {results/book_lsh200.csv};
    %\addlegendentry{Precision}
    \addplot[color=cyan,mark=*] table [x=Clusters, y=Recall, col sep=comma] {results/book_lsh200.csv};
    %\addlegendentry{Recall}
    \addplot[color=blue,mark=o] table [x=Clusters, y=F1-Score, col sep=comma] {results/book_lsh200.csv};
    %\addlegendentry{F1-Score}
    
    \end{axis}
    \end{tikzpicture}}
    \caption{EL 200}
    \label{fig:book_lsh200}

\end{subfigure}

\caption{Accuracy on library catalogue datasets.}
\label{fig:bk}
\end{figure}
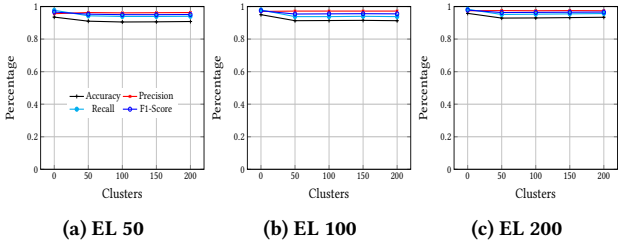
\end{multicols}

Fig.~\ref{fig:nc} demonstrates the accuracy evaluation on the NCVR dataset under different settings. Fig.~\ref{fig:nc} (a), (b) and (c) show the performance for the 10k dataset with encoding lengths 50, 100 and 200 and a varying number of clusters. It can be observed that the performance degrades significantly with length 50 after clustering. This is because a small length of encoding does not capture sufficient information, and results in poor clustering. With increased encoding length, we observe improvement in accuracy for both precision and recall. Thus, we suggest encoding length 200 for better performance with clustering.  %Similarly, Fig.~\ref{fig:nc} (d) and (e) show the performance for the 100k dataset with encoding lengths 100 and 200 and a varying number of clusters. Precision is not affected even after clustering. However, recall drops significantly with length 100 and marginally with 200. Thus, an accuracy tradeoff exists with clustering. Similar results can be observed in Fig.~\ref{fig:nc} (f) for the 1000k dataset with encoding length 200 and a varying number of clusters. %Also, it is important to note that, the precision is not impacted even with clustering, which is desirable in the context of finance.

Fig.~\ref{fig:bk} demonstrates the accuracy evaluation on the library catalogue dataset. The performance does not vary much with different encoding lengths. Encoding length 200 shows the best performance, with almost no difference in precision and recall. This gain is observed because of longer names in the book dataset, resulting in better representation even with smaller encoding lengths.

%\smallskip

%\textbf{Discussion:} We have only considered names for comparison in our fuzzy matching scheme. We also achieved high precision with the above datasets, choosing the right threshold. However, extra information like date of birth (dob) might be present, which can be used in case of high false positives. There are different ways to include extra information. Based on our experimental studies, we recommend concatenating LSH encoded dob (or any other extra information present and agreed by both parties) to LSH encoded name. However, this will increase the time and cost of search in the same ratio, similar to what we observed with increased LSH size for name encoding.

%\noindent\textbf{Discussion:} 
%More discussion on results and selection of parameters can be referred to in Appendix~\ref{sec-disc}.

\begin{table*}[!t]
\caption{Stepwise Computation and Communication Evaluation.}
\label{tab:nc-stepwise}
\resizebox{0.98\textwidth}{!}{
\begin{tabular}{|l|l|l|l|l|l|l|l|l|l|l|l|l|l|}
\hline
Step 1    & \multicolumn{2}{c}{Step 2 (\textit{A to B})}    &  & \multicolumn{2}{c}{Step 3 (\textit{B to A})} & &  \multicolumn{2}{c}{Step 4-5 (\textit{A to B})}  & & \multicolumn{2}{c}{Step 6-8 (\textit{B to A})} &  & Step 9    \\\hline

%PT (s) & PT (s) & \multicolumn{1}{c}{\textit{A to B}}       &           & PT (s) & \multicolumn{1}{c}{\textit{B to A}}       &           & PT (s) & \multicolumn{1}{c}{\textit{A to B}}         &           & PT (s) & \multicolumn{1}{c}{\textit{B to A}}       &           & PT (s) \\\hline

PT (s)          &  PT (s)         & CC (MB) & CT (s) &    PT (s)       & CC (MB) & CT (s) &    PT (s)       & CC (MB)   & CT (s) &     PT (s)      & CC (KB) & CT (s) &     PT (s)      \\\hline
6.08      & 0.562     & 89.1      & 0.01      & 0.968     & 15.7      & 0.001     & 0.144     & 22.3 + 22.3 & 0.002 & 0.26      & 175       & 0.0002    & 0.0003 \\\hline
\multicolumn{14}{l}{PT: Local Processing Time; CC and CT: Communication Cost and Time over LAN.}
\\
\hline
\end{tabular}}
\end{table*}

\begin{table*}[!t]
\centering
\caption{Computation and Communication Cost Analysis with varying settings.}
\label{tab:nc-cc}
\resizebox{0.98\textwidth}{!}{
\begin{tabular}{|p{1.65cm}|p{1.65cm}|p{1.65cm}|p{1.65cm}|p{1.65cm}|p{1.65cm}|p{1.65cm}|p{1.65cm}|p{1.65cm}|p{1.65cm}|p{1.65cm}|}
\hline
data size              & clusters & total cols & first round & time/col & memory & enc\_msg                 & enc\_msg               & sim\_cent & sign\_vec & comm/col                \\ \hline
                       &          &            & secs        & secs     & GB     & first round              & MB                     & MB        & MB        &                         \\ \hline
\multirow{5}{*}{10K}   & 0        &            &             & 97       & 0.4    &                          & \multirow{15}{*}{22.3} &           &           & 3.14 GB                 \\ \cline{2-7} \cline{9-11} 
                       & 50       & 308        & 1.53        & 0.26     & 6.63   & \multirow{4}{*}{89.1 MB} &                        & 15.7      & 22.3      & \multirow{4}{*}{175 KB} \\ \cline{2-6} \cline{9-10}
                       & 100      & 213        & 7.66        & 0.51     & 12.75  &                          &                        & 31.4      & 44.6      &                         \\ \cline{2-6} \cline{9-10}
                       & 150      & 154        & 15.6        & 0.75     & 18.88  &                          &                        & 47.1      & 66.9      &                         \\ \cline{2-6} \cline{9-10}
                       & 200      & 125        & 21.5        & 1        & 25     &                          &                        & 62.8      & 89.1      &                         \\ \cline{1-7} \cline{9-11} 
\multirow{5}{*}{100K}  & 0        &            &             & 986      &   1.2     &                          &                        &           &           & 31.4 GB                 \\ \cline{2-7} \cline{9-11} 
                       & 50       & 4206       & 1.53        & 0.26     & 6.63   & \multirow{4}{*}{89.1 MB} &                        & 15.7      & 22.3      & \multirow{4}{*}{175 KB} \\ \cline{2-6} \cline{9-10}
                       & 100      & 2544       & 7.66        & 0.51     & 12.75  &                          &                        & 31.4      & 44.6      &                         \\ \cline{2-6} \cline{9-10}
                       & 200      & 1769       & 21.5        & 1        & 25     &                          &                        & 62.8      & 89.1      &                         \\ \cline{2-6} \cline{9-10}
                       & 400      & 1173       & 229.22      & 2.21     & 50     &                          &                        & 126       & 178       &                         \\ \cline{1-7} \cline{9-11} 
\multirow{5}{*}{1000K} & 0        &            &             & 9945     &   6     &                          &                        &           &           & 314 GB                  \\ \cline{2-7} \cline{9-11} 
                       & 50       & 45,619     & 1.53        & 0.26     & 6.63   & \multirow{4}{*}{89.1 MB} &                        & 15.7      & 22.3      & \multirow{4}{*}{175 KB} \\ \cline{2-6} \cline{9-10}
                       & 100      & 37,310     & 7.66        & 0.51     & 12.75  &                          &                        & 31.4      & 44.6      &                         \\ \cline{2-6} \cline{9-10}
                       & 500      & 14,348     & 304.99      & 2.6      & 77.75  &                          &                        & 157       & 223       &                         \\ \cline{2-6} \cline{9-10}
                       & 1000     & 4,206      &     -        &    -      &   -     &                          &                        &   314        &           446 &                         \\ \hline
\multicolumn{11}{l}{first round (Step 2-3): time for matching with centroids; time/col (Step 6-8), repeated for col (column) i.e., length of the largest cluster;} \\ \multicolumn{11}{l}{enc\_msg: encrypted message of EL 200 and 50 (in Steps 2 and 5); sim\_cent: encrypted vector of similarity with centroids (in Step 3);} \\ \multicolumn{11}{l}{sign\_vec: encrypted sign vector mentioning most matching cluster (in Step 5); comm/col (Step 6-8), communication cost for each col.}
\\\hline
\end{tabular}}
\end{table*}

\noindent\textbf{Performance Evaluation.} 
Table~\ref{tab:nc-stepwise} presents the stepwise (Steps 1-9 in Section \ref{V.B}) computation and communication cost of executing our protocol. %It is evaluated over the North Carolina dataset. 
Organisation \textit{A} query size is 100 and \textit{B} stores 10K names. %Steps 1 to 9 are the same as discussed in Section \ref{V.B}. 
The number of clusters is set to 50. Step 1 is done offline, thus the time taken does not impact the overall search time. The latency of transferring ciphertext is negligible. Step 6–9 repeats for the length of the largest cluster. %The size of the ciphertexts depends on the encoding size and cluster size. It has been discussed below.
These results are reported over the NCVR dataset. However, most of these results are standard and not specific to a particular dataset; the ciphertext size depends on encoding length and the number of clusters. It is 89.1 MB in step 2, as the encoding size is 200, while it is 22.3 MB in step 5, with an encoding size of 50. The size of the sign vector in Step 5 is also 22.3 MB because, its size is 50, equal to the number of clusters.

Table~\ref{tab:nc-cc} presents an extensive evaluation, based on various dataset settings, encoding length, and number of clusters. It presents the time, memory, computation and communication costs. The number of queries is 100. Based on accuracy evaluation, it can be observed that an encoding length of 200 is more suitable for clustering. Therefore, we use a signature of length 200 for clustering, followed by matching with centroids. Further, matching with individual signatures of the best matching clusters is done over encoding length 50. This is because accuracy is similar with different encoding lengths without clustering, and operations over smaller encoding lengths reduce computation and communication costs.%, a crucial consideration in real-world applications.

If we cluster a dataset of size 10k into 50 clusters, the maximum size of a cluster is 308. To execute the scheme, the responder (\textit{B}) finds the cosine similarities of the received encrypted query with all the centroids. It takes 1.53 seconds to find cosine similarities with 50 centroids (of length 200). The size of the encrypted query in this round is 89.1 MB. This is the same throughout the experiment, irrespective of datasets,  since it is based on the encoding length of 200. 
Similarly, the size of the encrypted query for further matching is 22.3 MB with or without clustering, as it is based on the encoding length of 50. The size of encrypted cosine similarities with centroids from the first round is 15.7 MB, which is sent back to the querier (\textit{A}). Then, \textit{A} sends the encrypted sign values to \textit{B}, which is 22.3 MB. Both these sizes depend on the number of clusters. The time taken to perform the operation on each column is 0.26 seconds. This time involves operations of multiplying sign values to one column of names, adding them, and then finding cosine similarity. Then, \textit{B} returns the encrypted cosine similarity score to \textit{A} for each column. The size of the encrypted score is 175 KB for each column. 
Without clustering, communication in only sending back the result would have been 3.14 GB. 
Even in the worst scenario with 308 columns, it will reduce to 175KB $\times$ 308, i.e., 53.9 MB. The complete operation consumes 6.63 GB of memory. Increasing the number of clusters reduces the communication costs because of fewer columns. However, it has a higher memory requirement.

Similarly, if we cluster a dataset of 100k with 100 clusters, the maximum size of a cluster is 2544. The time taken to find the cosine similarity between the encrypted query and the centroids is 7.66 seconds. The time taken to compute each column is 0.51 seconds. The size of the encrypted score is 175 KB for each column. If no clustering had been used, the total communication involved in only sending back the result would have been 31.4 GB. 
Even in the worst scenario, with 2544 columns, it will reduce to 175 KB $\times$ 2544, i.e., 445.2 MB. When simulated on a single machine, the complete operation consumes 12.75 GB of memory. For a dataset of 1000k with 500 clusters, the maximum size of a cluster is 14348. If no clustering had been used, the total communication involved in only sending back the result would have been 314 GB. However, even in the worst scenario, with 14348 columns, it will reduce to 175 KB $\times$ 14348, i.e., 2.51~GB. The complete operation consumes 77.75 GB of memory when simulated on a single machine.  This demonstrates the reduction in communication cost that clustering offers. 
Due to higher memory requirements, it was not feasible to run experiments with 1000 clusters. In summary, it is evident that clustering not only reduces the communication cost significantly but also offers substantial computational gains. Next, we discuss the performance of clustering in terms of gain in computation.

%\smallskip

\begin{figure}[htbp!] 
\centering
\begin{subfigure}[b]{0.15\textwidth}
    \centering
    \resizebox{\textwidth}{\textwidth}{%
    \begin{tikzpicture}[thick, scale=1, every node/.style={transform shape}]
    \begin{axis}[
    xlabel={Columns},
    ylabel={Names},
    grid=both,
    legend style={at={(0.7,0.4)},anchor=north,draw={none}, font=\Large},
    label style={font=\huge},         ticklabel style = {font=\Large},
    ]
    
    % Array 1
    \addplot[mark=U,black] coordinates {(1,50) (2,100) (3,150) (4,200) (5,250) (6,300) (7,350) (8,400) (9,450) (10,500) (11,550) (12,600) (13,650) (14,700) (15,750) (16,800) (17,850) (18,900) (19,950) (20,1000) (21,1050) (22,1100) (23,1150) (24,1200) (25,1250) (26,1300) (27,1350) (28,1400) (29,1450) (30,1500) (31,1550) (32,1600) (33,1650) (34,1700) (35,1750) (36,1800) (37,1850) (38,1900) (39,1950) (40,2000) (41,2050) (42,2100) (43,2150) (44,2200) (45,2250) (46,2300) (47,2350) (48,2400) (49,2450) (50,2500) (51,2550) (52,2600) (53,2650) (54,2700) (55,2750) (56,2800) (57,2850) (58,2900) (59,2950) (60,3000) (61,3050) (62,3100) (63,3150) (64,3200) (65,3250) (66,3300) (67,3350) (68,3400) (69,3450) (70,3500) (71,3550) (72,3600) (73,3650) (74,3700) (75,3750) (76,3800) (77,3850) (78,3900) (79,3950) (80,4000) (81,4050) (82,4100) (83,4150) (84,4200) (85,4250) (86,4300) (87,4350) (88,4400) (89,4450) (90,4500) (91,4550) (92,4600) (93,4650) (94,4700) (95,4750) (96,4800) (97,4850) (98,4900) (99,4950) (100,5000) (101,5050) (102,5100) (103,5150) (104,5200) (105,5250) (106,5300) (107,5350) (108,5400) (109,5450) (110,5500) (111,5550) (112,5600) (113,5649) (114,5698) (115,5747) (116,5796) (117,5845) (118,5894) (119,5943) (120,5992) (121,6041) (122,6090) (123,6139) (124,6188) (125,6236) (126,6284) (127,6331) (128,6378) (129,6425) (130,6472) (131,6519) (132,6566) (133,6612) (134,6658) (135,6704) (136,6749) (137,6794) (138,6839) (139,6884) (140,6929) (141,6974) (142,7019) (143,7064) (144,7106) (145,7148) (146,7190) (147,7232) (148,7274) (149,7315) (150,7356) (151,7395) (152,7434) (153,7473) (154,7512) (155,7551) (156,7590) (157,7629) (158,7668) (159,7707) (160,7746) (161,7785) (162,7823) (163,7861) (164,7898) (165,7935) (166,7972) (167,8008) (168,8043) (169,8077) (170,8111) (171,8145) (172,8179) (173,8213) (174,8247) (175,8281) (176,8315) (177,8349) (178,8381) (179,8412) (180,8443) (181,8474) (182,8504) (183,8533) (184,8562) (185,8590) (186,8618) (187,8646) (188,8674) (189,8702) (190,8729) (191,8755) (192,8779) (193,8803) (194,8826) (195,8849) (196,8870) (197,8891) (198,8912) (199,8933) (200,8953) (201,8973) (202,8992) (203,9011) (204,9030) (205,9049) (206,9067) (207,9085) (208,9103) (209,9121) (210,9139) (211,9157) (212,9175) (213,9193) (214,9211) (215,9229) (216,9247) (217,9265) (218,9282) (219,9299) (220,9315) (221,9331) (222,9347) (223,9363) (224,9379) (225,9395) (226,9411) (227,9427) (228,9443) (229,9459) (230,9474) (231,9489) (232,9504) (233,9519) (234,9532) (235,9545) (236,9558) (237,9570) (238,9582) (239,9593) (240,9604) (241,9614) (242,9624) (243,9634) (244,9644) (245,9654) (246,9664) (247,9674) (248,9683) (249,9692) (250,9701) (251,9709) (252,9717) (253,9725) (254,9733) (255,9741) (256,9748) (257,9755) (258,9762) (259,9769) (260,9776) (261,9783) (262,9790) (263,9797) (264,9804) (265,9811) (266,9818) (267,9825) (268,9832) (269,9839) (270,9846) (271,9853) (272,9859) (273,9865) (274,9871) (275,9877) (276,9883) (277,9889) (278,9895) (279,9901) (280,9907) (281,9913) (282,9918) (283,9923) (284,9928) (285,9933) (286,9938) (287,9943) (288,9948) (289,9953) (290,9958) (291,9962) (292,9966) (293,9969) (294,9972) (295,9975) (296,9978) (297,9981) (298,9984) (299,9986) (300,9988) (301,9990) (302,9992) (303,9994) (304,9996) (305,9997) (306,9998) (307,9999) (308,10000)};
    \addlegendentry{Clusters 50}
    
    % Array 2
    \addplot[mark=U,red] coordinates {(1,100) (2,200) (3,300) (4,400) (5,500) (6,600) (7,700) (8,800) (9,900) (10,1000) (11,1100) (12,1200) (13,1300) (14,1400) (15,1500) (16,1600) (17,1700) (18,1800) (19,1900) (20,2000) (21,2100) (22,2200) (23,2300) (24,2400) (25,2500) (26,2600) (27,2700) (28,2800) (29,2900) (30,3000) (31,3100) (32,3200) (33,3300) (34,3400) (35,3500) (36,3600) (37,3699) (38,3798) (39,3897) (40,3995) (41,4093) (42,4191) (43,4289) (44,4386) (45,4483) (46,4580) (47,4677) (48,4774) (49,4871) (50,4967) (51,5063) (52,5158) (53,5252) (54,5345) (55,5438) (56,5531) (57,5623) (58,5714) (59,5805) (60,5896) (61,5986) (62,6076) (63,6166) (64,6251) (65,6335) (66,6418) (67,6500) (68,6582) (69,6663) (70,6742) (71,6821) (72,6897) (73,6972) (74,7045) (75,7117) (76,7188) (77,7257) (78,7325) (79,7391) (80,7455) (81,7518) (82,7579) (83,7638) (84,7697) (85,7755) (86,7813) (87,7871) (88,7926) (89,7981) (90,8032) (91,8082) (92,8131) (93,8178) (94,8223) (95,8266) (96,8308) (97,8348) (98,8388) (99,8428) (100,8468) (101,8508) (102,8547) (103,8584) (104,8619) (105,8650) (106,8681) (107,8712) (108,8743) (109,8774) (110,8805) (111,8836) (112,8866) (113,8895) (114,8924) (115,8953) (116,8982) (117,9010) (118,9037) (119,9063) (120,9089) (121,9115) (122,9141) (123,9166) (124,9191) (125,9216) (126,9241) (127,9266) (128,9290) (129,9314) (130,9337) (131,9359) (132,9380) (133,9401) (134,9422) (135,9441) (136,9459) (137,9477) (138,9494) (139,9511) (140,9526) (141,9541) (142,9556) (143,9570) (144,9584) (145,9598) (146,9611) (147,9624) (148,9637) (149,9650) (150,9663) (151,9675) (152,9686) (153,9697) (154,9708) (155,9719) (156,9729) (157,9738) (158,9747) (159,9756) (160,9764) (161,9772) (162,9780) (163,9788) (164,9796) (165,9804) (166,9812) (167,9820) (168,9828) (169,9836) (170,9844) (171,9852) (172,9859) (173,9865) (174,9871) (175,9877) (176,9883) (177,9889) (178,9895) (179,9901) (180,9907) (181,9913) (182,9919) (183,9925) (184,9930) (185,9935) (186,9940) (187,9945) (188,9950) (189,9954) (190,9958) (191,9961) (192,9964) (193,9967) (194,9970) (195,9973) (196,9976) (197,9978) (198,9980) (199,9982) (200,9984) (201,9986) (202,9988) (203,9990) (204,9991) (205,9992) (206,9993) (207,9994) (208,9995) (209,9996) (210,9997) (211,9998) (212,9999) (213,10000)};
    \addlegendentry{Clusters 100}
    
    % Array 3
    \addplot[mark=U,cyan] coordinates {(1,150) (2,300) (3,450) (4,600) (5,750) (6,900) (7,1050) (8,1200) (9,1350) (10,1500) (11,1650) (12,1800) (13,1950) (14,2100) (15,2250) (16,2400) (17,2550) (18,2700) (19,2849) (20,2998) (21,3147) (22,3296) (23,3445) (24,3594) (25,3743) (26,3892) (27,4041) (28,4189) (29,4336) (30,4482) (31,4626) (32,4770) (33,4912) (34,5051) (35,5188) (36,5324) (37,5458) (38,5591) (39,5722) (40,5849) (41,5973) (42,6096) (43,6218) (44,6336) (45,6452) (46,6567) (47,6680) (48,6791) (49,6897) (50,7002) (51,7103) (52,7203) (53,7299) (54,7389) (55,7476) (56,7562) (57,7646) (58,7727) (59,7806) (60,7883) (61,7959) (62,8027) (63,8095) (64,8162) (65,8227) (66,8290) (67,8352) (68,8411) (69,8468) (70,8523) (71,8577) (72,8629) (73,8679) (74,8727) (75,8774) (76,8819) (77,8860) (78,8899) (79,8937) (80,8975) (81,9012) (82,9049) (83,9085) (84,9119) (85,9150) (86,9181) (87,9212) (88,9242) (89,9271) (90,9299) (91,9326) (92,9352) (93,9378) (94,9404) (95,9430) (96,9454) (97,9478) (98,9501) (99,9524) (100,9547) (101,9567) (102,9586) (103,9605) (104,9623) (105,9640) (106,9656) (107,9672) (108,9687) (109,9702) (110,9717) (111,9732) (112,9747) (113,9762) (114,9776) (115,9788) (116,9799) (117,9810) (118,9821) (119,9831) (120,9841) (121,9851) (122,9860) (123,9869) (124,9878) (125,9887) (126,9895) (127,9903) (128,9911) (129,9918) (130,9925) (131,9931) (132,9936) (133,9941) (134,9946) (135,9951) (136,9956) (137,9961) (138,9966) (139,9970) (140,9974) (141,9978) (142,9982) (143,9984) (144,9986) (145,9988) (146,9990) (147,9992) (148,9994) (149,9995) (150,9996) (151,9997) (152,9998) (153,9999) (154,10000)};
    \addlegendentry{Clusters 150}

    % Array 4
    \addplot[mark=U,blue] coordinates {(1,200) (2,400) (3,600) (4,800) (5,1000) (6,1200) (7,1400) (8,1600) (9,1800) (10,2000) (11,2200) (12,2400) (13,2600) (14,2800) (15,3000) (16,3200) (17,3400) (18,3599) (19,3798) (20,3997) (21,4195) (22,4393) (23,4591) (24,4787) (25,4978) (26,5166) (27,5350) (28,5523) (29,5694) (30,5862) (31,6026) (32,6188) (33,6347) (34,6500) (35,6649) (36,6793) (37,6935) (38,7072) (39,7202) (40,7331) (41,7452) (42,7567) (43,7674) (44,7776) (45,7876) (46,7971) (47,8062) (48,8150) (49,8235) (50,8319) (51,8400) (52,8480) (53,8558) (54,8632) (55,8703) (56,8773) (57,8840) (58,8904) (59,8962) (60,9016) (61,9066) (62,9113) (63,9159) (64,9205) (65,9249) (66,9288) (67,9325) (68,9360) (69,9393) (70,9423) (71,9452) (72,9480) (73,9505) (74,9530) (75,9554) (76,9576) (77,9598) (78,9620) (79,9642) (80,9663) (81,9684) (82,9705) (83,9725) (84,9745) (85,9763) (86,9780) (87,9796) (88,9810) (89,9823) (90,9836) (91,9849) (92,9860) (93,9871) (94,9882) (95,9892) (96,9902) (97,9912) (98,9921) (99,9929) (100,9937) (101,9944) (102,9950) (103,9956) (104,9962) (105,9968) (106,9973) (107,9978) (108,9981) (109,9983) (110,9985) (111,9986) (112,9987) (113,9988) (114,9989) (115,9990) (116,9991) (117,9992) (118,9993) (119,9994) (120,9995) (121,9996) (122,9997) (123,9998) (124,9999) (125,10000)};
    \addlegendentry{Clusters 200}
    
    \end{axis}
    \end{tikzpicture}}
    \caption{10K}
    \label{fig:nc10k_lsh200_c}
\end{subfigure}
\begin{subfigure}[b]{0.15\textwidth}
    \centering
    \resizebox{\textwidth}{\textwidth}{%
    \begin{tikzpicture}[thick, scale=1, every node/.style={transform shape}]
    \begin{axis}[
    xlabel={Columns},
    ylabel={Names},
    grid=both,
    legend style={at={(0.7,0.4)},anchor=north,draw={none}, font=\Large},
    label style={font=\huge},         ticklabel style = {font=\Large},
    ]
    
    % Array 1
    \addplot[mark=U,black] coordinates {(1,50) (101,5050) (201,10050) (301,15050) (401,20050) (501,25050) (601,30050) (701,35050) (801,40050) (901,45050) (1001,50050) (1101,55050) (1201,60030) (1301,64751) (1401,69005) (1501,72681) (1601,75992) (1701,79291) (1801,82380) (1901,84926) (2001,86921) (2101,88722) (2201,90418) (2301,91858) (2401,93210) (2501,94500) (2601,95623) (2701,96514) (2801,97268) (2901,97723) (3001,98023) (3101,98323) (3201,98623) (3301,98864) (3401,99064) (3501,99264) (3601,99395) (3701,99495) (3801,99595) (3901,99695) (4001,99795) (4101,99895) (4201,99995)};
    \addlegendentry{Clusters 50}
    
    % Array 2
    \addplot[mark=U,red] coordinates {(1,100) (101,10100) (201,20100) (301,30100) (401,40100) (501,50003) (601,59216) (701,67322) (801,74257) (901,79484) (1001,83602) (1101,86911) (1201,89836) (1301,92235) (1401,93900) (1501,95178) (1601,96296) (1701,97188) (1801,97860) (1901,98376) (2001,98692) (2101,98992) (2201,99292) (2301,99592) (2401,99857) (2501,99957)};
    \addlegendentry{Clusters 100}
    
    % Array 3
    \addplot[mark=U,cyan] coordinates {(1,200) (101,20040) (201,39558) (301,58097) (401,72653) (501,82421) (601,87934) (701,91477) (801,94327) (901,96418) (1001,97873) (1101,98759) (1201,99133) (1301,99333) (1401,99533) (1501,99732) (1601,99832) (1701,99932)};
    \addlegendentry{Clusters 200}
    
    % Array 4
    \addplot[mark=U,blue] coordinates {(1,400) (101,39418) (201,70798) (301,87406) (401,94575) (501,97048) (601,98477) (701,99151) (801,99470) (901,99670) (1001,99828) (1101,99928)};
    \addlegendentry{Clusters 400}
    
    \end{axis}
    \end{tikzpicture}}
    \caption{100K}
    \label{fig:nc100k_lsh200_c}
\end{subfigure}
\begin{subfigure}[b]{0.15\textwidth}
    \centering
    \resizebox{\textwidth}{\textwidth}{%
    \begin{tikzpicture}[thick, scale=1, every node/.style={transform shape}]
    \begin{axis}[
    xlabel={Columns},
    ylabel={Names},
    grid=both,
    legend style={at={(0.7,0.4)},anchor=north,draw={none}, font=\Large},
    label style={font=\huge},         ticklabel style = {font=\Large},
    ]

    % Array 1
    \addplot[mark=U,black] coordinates {(1,50) (101,5050) (201,10050) (301,15050) (401,20050) (501,25050) (601,30050) (701,35050) (801,40050) (901,45050) (1001,50050) (1101,55050) (1201,60050) (1301,65050) (1401,70050) (1501,75050) (1601,80050) (1701,85050) (1801,90050) (1901,95050) (2001,100050) (2101,105050) (2201,110050) (2301,115050) (2401,120050) (2501,125050) (2601,130050) (2701,135050) (2801,140050) (2901,145050) (3001,150050) (3101,155050) (3201,160050) (3301,165050) (3401,170050) (3501,175050) (3601,180050) (3701,185050) (3801,190050) (3901,195050) (4001,200050) (4101,205050) (4201,210050) (4301,215050) (4401,220050) (4501,225050) (4601,230050) (4701,235050) (4801,240050) (4901,245050) (5001,250050) (5101,255050) (5201,260050) (5301,265050) (5401,270050) (5501,275050) (5601,280050) (5701,285050) (5801,290050) (5901,295050) (6001,300050) (6101,305050) (6201,310050) (6301,315050) (6401,320050) (6501,325050) (6601,330050) (6701,335050) (6801,340050) (6901,345050) (7001,350050) (7101,355050) (7201,360050) (7301,365050) (7401,370050) (7501,375050) (7601,380050) (7701,385050) (7801,390050) (7901,395050) (8001,400050) (8101,405050) (8201,410050) (8301,415050) (8401,420050) (8501,425050) (8601,430050) (8701,435050) (8801,440050) (8901,445050) (9001,450050) (9101,455050) (9201,460050) (9301,465050) (9401,470050) (9501,475050) (9601,480050) (9701,485050) (9801,490050) (9901,495050) (10001,500050) (10101,505050) (10201,510050) (10301,515050) (10401,520050) (10501,525050) (10601,530050) (10701,535050) (10801,540050) (10901,545050) (11001,550050) (11101,555050) (11201,560050) (11301,565050) (11401,570023) (11501,574840) (11601,579640) (11701,584440) (11801,589240) (11901,594040) (12001,598840) (12101,603640) (12201,608440) (12301,613182) (12401,617882) (12501,622582) (12601,627282) (12701,631982) (12801,636682) (12901,641284) (13001,645849) (13101,650349) (13201,654849) (13301,659272) (13401,663672) (13501,668072) (13601,672404) (13701,676704) (13801,681004) (13901,685256) (14001,689361) (14101,693307) (14201,697207) (14301,701107) (14401,704922) (14501,708622) (14601,712293) (14701,715826) (14801,719326) (14901,722826) (15001,726326) (15101,729826) (15201,733273) (15301,736666) (15401,739867) (15501,743067) (15601,746178) (15701,749278) (15801,752378) (15901,755478) (16001,758578) (16101,761659) (16201,764659) (16301,767659) (16401,770659) (16501,773564) (16601,776464) (16701,779361) (16801,782161) (16901,784961) (17001,787761) (17101,790505) (17201,793174) (17301,795774) (17401,798339) (17501,800839) (17601,803339) (17701,805839) (17801,808339) (17901,810839) (18001,813339) (18101,815839) (18201,818339) (18301,820839) (18401,823339) (18501,825839) (18601,828339) (18701,830839) (18801,833314) (18901,835714) (19001,838114) (19101,840514) (19201,842855) (19301,845155) (19401,847414) (19501,849614) (19601,851814) (19701,854014) (19801,856159) (19901,858186) (20001,860186) (20101,862186) (20201,864186) (20301,866186) (20401,868186) (20501,870186) (20601,872186) (20701,874186) (20801,876186) (20901,878186) (21001,880186) (21101,882100) (21201,883900) (21301,885700) (21401,887500) (21501,889300) (21601,891100) (21701,892900) (21801,894700) (21901,896461) (22001,898161) (22101,899861) (22201,901550) (22301,903150) (22401,904737) (22501,906237) (22601,907694) (22701,909094) (22801,910494) (22901,911894) (23001,913294) (23101,914694) (23201,916092) (23301,917392) (23401,918692) (23501,919992) (23601,921292) (23701,922592) (23801,923892) (23901,925192) (24001,926492) (24101,927744) (24201,928944) (24301,930144) (24401,931344) (24501,932544) (24601,933744) (24701,934944) (24801,936144) (24901,937344) (25001,938458) (25101,939558) (25201,940658) (25301,941758) (25401,942858) (25501,943958) (25601,945058) (25701,946158) (25801,947255) (25901,948255) (26001,949255) (26101,950255) (26201,951255) (26301,952255) (26401,953255) (26501,954255) (26601,955255) (26701,956255) (26801,957255) (26901,958255) (27001,959255) (27101,960190) (27201,960990) (27301,961790) (27401,962590) (27501,963390) (27601,964190) (27701,964990) (27801,965790) (27901,966576) (28001,967276) (28101,967976) (28201,968676) (28301,969376) (28401,970076) (28501,970708) (28601,971308) (28701,971908) (28801,972508) (28901,973108) (29001,973708) (29101,974308) (29201,974908) (29301,975508) (29401,976055) (29501,976555) (29601,977055) (29701,977555) (29801,977975) (29901,978375) (30001,978775) (30101,979175) (30201,979575) (30301,979975) (30401,980375) (30501,980775) (30601,981175) (30701,981575) (30801,981975) (30901,982375) (31001,982775) (31101,983096) (31201,983396) (31301,983696) (31401,983965) (31501,984165) (31601,984365) (31701,984565) (31801,984765) (31901,984965) (32001,985165) (32101,985365) (32201,985565) (32301,985765) (32401,985965) (32501,986165) (32601,986365) (32701,986565) (32801,986765) (32901,986965) (33001,987165) (33101,987365) (33201,987565) (33301,987682) (33401,987782) (33501,987882) (33601,987982) (33701,988082) (33801,988182) (33901,988282) (34001,988382) (34101,988482) (34201,988582) (34301,988682) (34401,988782) (34501,988882) (34601,988982) (34701,989082) (34801,989182) (34901,989282) (35001,989382) (35101,989482) (35201,989582) (35301,989682) (35401,989782) (35501,989882) (35601,989982) (35701,990082) (35801,990182) (35901,990282) (36001,990382) (36101,990482) (36201,990582) (36301,990682) (36401,990782) (36501,990882) (36601,990982) (36701,991082) (36801,991182) (36901,991282) (37001,991382) (37101,991482) (37201,991582) (37301,991682) (37401,991782) (37501,991882) (37601,991982) (37701,992082) (37801,992182) (37901,992282) (38001,992382) (38101,992482) (38201,992582) (38301,992682) (38401,992782) (38501,992882) (38601,992982) (38701,993082) (38801,993182) (38901,993282) (39001,993382) (39101,993482) (39201,993582) (39301,993682) (39401,993782) (39501,993882) (39601,993982) (39701,994082) (39801,994182) (39901,994282) (40001,994382) (40101,994482) (40201,994582) (40301,994682) (40401,994782) (40501,994882) (40601,994982) (40701,995082) (40801,995182) (40901,995282) (41001,995382) (41101,995482) (41201,995582) (41301,995682) (41401,995782) (41501,995882) (41601,995982) (41701,996082) (41801,996182) (41901,996282) (42001,996382) (42101,996482) (42201,996582) (42301,996682) (42401,996782) (42501,996882) (42601,996982) (42701,997082) (42801,997182) (42901,997282) (43001,997382) (43101,997482) (43201,997582) (43301,997682) (43401,997782) (43501,997882) (43601,997982) (43701,998082) (43801,998182) (43901,998282) (44001,998382) (44101,998482) (44201,998582) (44301,998682) (44401,998782) (44501,998882) (44601,998982) (44701,999082) (44801,999182) (44901,999282) (45001,999382) (45101,999482) (45201,999582) (45301,999682) (45401,999782) (45501,999882) (45601,999982)};
    \addlegendentry{Clusters 50}

    % Array 2
    \addplot[mark=U,red] coordinates {(1,100) (101,10100) (201,20100) (301,30100) (401,40100) (501,50100) (601,60100) (701,70100) (801,80100) (901,90100) (1001,100100) (1101,110100) (1201,120100) (1301,130100) (1401,140100) (1501,150100) (1601,160100) (1701,170100) (1801,180100) (1901,190100) (2001,200100) (2101,210100) (2201,220100) (2301,230100) (2401,240100) (2501,250100) (2601,260100) (2701,270100) (2801,280100) (2901,290100) (3001,300100) (3101,310100) (3201,320100) (3301,330100) (3401,340039) (3501,349939) (3601,359839) (3701,369739) (3801,379639) (3901,389539) (4001,399439) (4101,409262) (4201,419062) (4301,428715) (4401,438315) (4501,447915) (4601,457499) (4701,466999) (4801,476499) (4901,485999) (5001,495490) (5101,504890) (5201,514290) (5301,523690) (5401,533090) (5501,542439) (5601,551672) (5701,560872) (5801,570003) (5901,578962) (6001,587862) (6101,596762) (6201,605589) (6301,614373) (6401,622868) (6501,631268) (6601,639668) (6701,648068) (6801,656323) (6901,664423) (7001,672498) (7101,680398) (7201,688221) (7301,695782) (7401,702985) (7501,709985) (7601,716856) (7701,723604) (7801,730165) (7901,736404) (8001,742504) (8101,748524) (8201,754507) (8301,760144) (8401,765356) (8501,770556) (8601,775659) (8701,780544) (8801,785328) (8901,789891) (9001,794130) (9101,798186) (9201,802157) (9301,805922) (9401,809525) (9501,813125) (9601,816640) (9701,819940) (9801,823240) (9901,826540) (10001,829751) (10101,832924) (10201,835954) (10301,838954) (10401,841854) (10501,844734) (10601,847534) (10701,850334) (10801,853062) (10901,855762) (11001,858462) (11101,861130) (11201,863678) (11301,866178) (11401,868678) (11501,871178) (11601,873678) (11701,876144) (11801,878544) (11901,880944) (12001,883344) (12101,885744) (12201,888093) (12301,890393) (12401,892693) (12501,894809) (12601,896749) (12701,898649) (12801,900435) (12901,902135) (13001,903835) (13101,905535) (13201,907235) (13301,908920) (13401,910470) (13501,911970) (13601,913470) (13701,914957) (13801,916357) (13901,917757) (14001,919157) (14101,920557) (14201,921957) (14301,923328) (14401,924628) (14501,925928) (14601,927189) (14701,928383) (14801,929483) (14901,930583) (15001,931683) (15101,932724) (15201,933724) (15301,934724) (15401,935724) (15501,936724) (15601,937724) (15701,938724) (15801,939724) (15901,940647) (16001,941547) (16101,942447) (16201,943347) (16301,944247) (16401,945147) (16501,946047) (16601,946947) (16701,947847) (16801,948747) (16901,949647) (17001,950547) (17101,951447) (17201,952347) (17301,953247) (17401,954147) (17501,955047) (17601,955947) (17701,956847) (17801,957747) (17901,958647) (18001,959547) (18101,960447) (18201,961347) (18301,962247) (18401,963060) (18501,963860) (18601,964660) (18701,965460) (18801,966260) (18901,967060) (19001,967860) (19101,968660) (19201,969460) (19301,970260) (19401,971060) (19501,971742) (19601,972342) (19701,972942) (19801,973542) (19901,974142) (20001,974742) (20101,975342) (20201,975942) (20301,976489) (20401,976989) (20501,977489) (20601,977919) (20701,978319) (20801,978719) (20901,979119) (21001,979519) (21101,979919) (21201,980281) (21301,980581) (21401,980881) (21501,981181) (21601,981481) (21701,981781) (21801,982081) (21901,982381) (22001,982655) (22101,982855) (22201,983055) (22301,983255) (22401,983455) (22501,983655) (22601,983855) (22701,984055) (22801,984255) (22901,984455) (23001,984655) (23101,984855) (23201,985055) (23301,985255) (23401,985455) (23501,985655) (23601,985855) (23701,986055) (23801,986255) (23901,986455) (24001,986655) (24101,986791) (24201,986891) (24301,986991) (24401,987091) (24501,987191) (24601,987291) (24701,987391) (24801,987491) (24901,987591) (25001,987691) (25101,987791) (25201,987891) (25301,987991) (25401,988091) (25501,988191) (25601,988291) (25701,988391) (25801,988491) (25901,988591) (26001,988691) (26101,988791) (26201,988891) (26301,988991) (26401,989091) (26501,989191) (26601,989291) (26701,989391) (26801,989491) (26901,989591) (27001,989691) (27101,989791) (27201,989891) (27301,989991) (27401,990091) (27501,990191) (27601,990291) (27701,990391) (27801,990491) (27901,990591) (28001,990691) (28101,990791) (28201,990891) (28301,990991) (28401,991091) (28501,991191) (28601,991291) (28701,991391) (28801,991491) (28901,991591) (29001,991691) (29101,991791) (29201,991891) (29301,991991) (29401,992091) (29501,992191) (29601,992291) (29701,992391) (29801,992491) (29901,992591) (30001,992691) (30101,992791) (30201,992891) (30301,992991) (30401,993091) (30501,993191) (30601,993291) (30701,993391) (30801,993491) (30901,993591) (31001,993691) (31101,993791) (31201,993891) (31301,993991) (31401,994091) (31501,994191) (31601,994291) (31701,994391) (31801,994491) (31901,994591) (32001,994691) (32101,994791) (32201,994891) (32301,994991) (32401,995091) (32501,995191) (32601,995291) (32701,995391) (32801,995491) (32901,995591) (33001,995691) (33101,995791) (33201,995891) (33301,995991) (33401,996091) (33501,996191) (33601,996291) (33701,996391) (33801,996491) (33901,996591) (34001,996691) (34101,996791) (34201,996891) (34301,996991) (34401,997091) (34501,997191) (34601,997291) (34701,997391) (34801,997491) (34901,997591) (35001,997691) (35101,997791) (35201,997891) (35301,997991) (35401,998091) (35501,998191) (35601,998291) (35701,998391) (35801,998491) (35901,998591) (36001,998691) (36101,998791) (36201,998891) (36301,998991) (36401,999091) (36501,999191) (36601,999291) (36701,999391) (36801,999491) (36901,999591) (37001,999691) (37101,999791) (37201,999891) (37301,999991)};
    \addlegendentry{Clusters 100}

    % Array 3
    \addplot[mark=U,cyan] coordinates {(1,500) (101,50500) (201,100500) (301,150500) (401,200478) (501,250208) (601,299453) (701,347691) (801,394285) (901,438771) (1001,480327) (1101,519527) (1201,556090) (1301,589748) (1401,620633) (1501,649420) (1601,676384) (1701,701088) (1801,723698) (1901,744272) (2001,762674) (2101,779551) (2201,794494) (2301,808128) (2401,820851) (2501,832435) (2601,842939) (2701,852586) (2801,861207) (2901,869005) (3001,876456) (3101,883503) (3201,890160) (3301,896383) (3401,901988) (3501,907073) (3601,911738) (3701,916138) (3801,920379) (3901,924306) (4001,927967) (4101,931431) (4201,934590) (4301,937590) (4401,940529) (4501,943207) (4601,945517) (4701,947688) (4801,949729) (4901,951650) (5001,953335) (5101,954935) (5201,956532) (5301,958032) (5401,959452) (5501,960752) (5601,962052) (5701,963352) (5801,964652) (5901,965952) (6001,967167) (6101,968210) (6201,969210) (6301,970210) (6401,971210) (6501,972210) (6601,973210) (6701,974210) (6801,975210) (6901,976190) (7001,977024) (7101,977731) (7201,978431) (7301,979131) (7401,979831) (7501,980531) (7601,981231) (7701,981931) (7801,982631) (7901,983331) (8001,984013) (8101,984613) (8201,985213) (8301,985813) (8401,986413) (8501,987013) (8601,987588) (8701,988088) (8801,988588) (8901,989088) (9001,989588) (9101,990088) (9201,990588) (9301,991088) (9401,991571) (9501,991971) (9601,992359) (9701,992659) (9801,992959) (9901,993259) (10001,993559) (10101,993859) (10201,994159) (10301,994459) (10401,994759) (10501,995059) (10601,995359) (10701,995659) (10801,995864) (10901,996064) (11001,996264) (11101,996464) (11201,996664) (11301,996864) (11401,997053) (11501,997153) (11601,997253) (11701,997353) (11801,997453) (11901,997553) (12001,997653) (12101,997753) (12201,997853) (12301,997953) (12401,998053) (12501,998153) (12601,998253) (12701,998353) (12801,998453) (12901,998553) (13001,998653) (13101,998753) (13201,998853) (13301,998953) (13401,999053) (13501,999153) (13601,999253) (13701,999353) (13801,999453) (13901,999553) (14001,999653) (14101,999753) (14201,999853) (14301,999953)};
    \addlegendentry{Clusters 500}

    % Array 4
    \addplot[mark=U,blue] coordinates {(1,1000) (101,100966) (201,200672) (301,298256) (401,390720) (501,474446) (601,549250) (701,613012) (801,666153) (901,710757) (1001,748570) (1101,781440) (1201,809672) (1301,834163) (1401,855704) (1501,873549) (1601,888426) (1701,901149) (1801,911600) (1901,920793) (2001,928980) (2101,936186) (2201,942555) (2301,947950) (2401,952605) (2501,956598) (2601,960044) (2701,962874) (2801,965275) (2901,967451) (3001,969551) (3101,971552) (3201,973414) (3301,975024) (3401,976371) (3501,977671) (3601,978850) (3701,979950) (3801,981050) (3901,982150) (4001,983250) (4101,984330) (4201,985330) (4301,986242) (4401,987042) (4501,987765) (4601,988465) (4701,989097) (4801,989697) (4901,990297) (5001,990897) (5101,991497) (5201,992097) (5301,992697) (5401,993297) (5501,993810) (5601,994278) (5701,994668) (5801,994968) (5901,995268) (6001,995568) (6101,995868) (6201,996168) (6301,996468) (6401,996768) (6501,997068) (6601,997368) (6701,997668) (6801,997902) (6901,998102) (7001,998302) (7101,998502) (7201,998702) (7301,998902) (7401,999102) (7501,999271) (7601,999371) (7701,999471) (7801,999571) (7901,999671) (8001,999771) (8101,999871) (8201,999971)};
    \addlegendentry{Clusters 1000}
    
    % Array 2
    %\addplot[mark=x,red] coordinates { };
    %\addlegendentry{Clusters 5000}

    % Array 1
    %\addplot[mark=+,red] table[x index=0, y index=1] {};
    %\addlegendentry{Array 1}
    
    % Array 2
    %\addplot[mark=x,blue] table[x index=0, y index=1] {\ncfivetk};
    %\addlegendentry{Array 2}

    \end{axis}
    \end{tikzpicture}}
    \caption{1000K}
    \label{fig:nc1000k_lsh200_c}
\end{subfigure}
\caption{Clustering coverage on NCVR dataset.}
\label{fig:nc_c}
\end{figure}
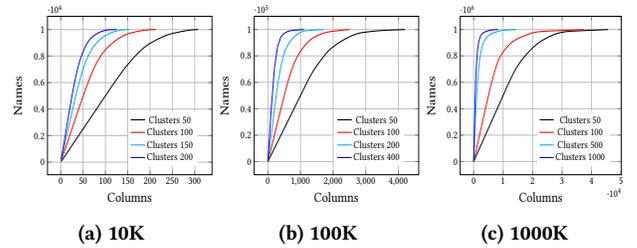

%Due to the maximum number of elements in a cluster, the time to search could grow very long. However, this is the worst-case scenario due to higher numbers of names in a particular cluster. 
Fig.~\ref{fig:nc_c} demonstrates clustering done on the NCVR dataset for varying numbers of clusters over different data sizes. As shown in Table~\ref{tab:nc-cc}, for a dataset of 100K records, the time taken for linear search with 100 queries is 986 seconds. With clustering, it can be observed in Fig.~\ref{fig:nc_c} (b) that though the maximum number of columns with 50 clusters is 4206, more than 50\% of names are covered in 1000 columns, 80\% in 2000 columns and 90\% in 2500 columns. As the time taken to compute on each column in the given settings is 0.26 seconds, it would take a maximum of 260 seconds to search from 50K names, 520 seconds from 80K names, and 650 seconds from 90K names. %Thus, it improves the search time for most cases.

Similarly, for a dataset of 1000K, it takes 9945 seconds to search 100 queries linearly. With clustering, the maximum number of columns with 100 clusters is 37,310, but around 80\% of names are covered only in 10000 and 90\% in 15000 columns, as shown in Fig.~\ref{fig:nc_c}~(c).
As the time taken to perform the operation on each column in the given settings is 0.51 seconds, it would take a maximum of 5100 and 7650 seconds to search from 800K and 900K names, respectively. It shows a significant gain in search time compared to linear search for most cases. Increasing the number of clusters increases the time for the first round and operation per column; however, a large number of clusters results in better coverage. Consequently, the maximum number of searches is covered in fewer columns. Hence, choosing the number of clusters influences the performance.
%\smallskip

\noindent\textbf{Comparison.} We compare our scheme in terms of security, accuracy and performance to similar works in~\cite{essex2019secure,khurram2020sfour,weicryptographically}. 

\noindent
\textit{Security:} The secure approximate string matching proposed in~\cite{essex2019secure} is based on additive homomorphic encryption schemes. 
The security of such schemes is typically based on well-studied mathematical assumptions specific to the homomorphic property they support.
Because they simplify certain aspects of HE, they also bring a trade-off with security. 
For example, the security of the Paillier cryptosystem is based on the Decisional Composite Residuosity Assumption. FHE schemes often rely on more advanced mathematical structures, such as lattice-based cryptography, which provides a higher level of security. SMPC-based schemes~\cite{khurram2020sfour,weicryptographically} performed private record linkage. These schemes make both parties learn about the fuzzy private set intersection, which does not fulfill the requirements for financial regulations. Moreover, they reveal the number of comparisons, which also reveal sensitive information to both parties. Thus, these approaches are not suitable from the regulatory and security perspectives.

\begin{comment}
\begin{table}[h]
\caption{Accuracy comparison on NCVR dataset.}
\label{tab:comp}
\centering
%\resizebox{0.48\textwidth}{!}{
\begin{tabular}{|c|c|c|c|c|}
\hline
SFour~\cite{khurram2020sfour}  & SFour (Best) & SPRL~\cite{weicryptographically} & Ours & Ours (Best) \\ \hline
97.9 & 98.6  & 99.97                                 & 96.1              & 98.6        \\ \hline
\end{tabular}%}
\end{table}
\end{comment}

\noindent
\textit{Accuracy comparison:} The HE-based solution in {\cite{essex2019secure} discusses the effectiveness of the scheme in terms of reducing false positives. It achieves a negligible false positive rate with a match threshold of 0.8, based on their scheme. Similarly, our approach achieved a high precision of 99.5\% with a matching threshold of 0.8, as shown in Fig.~\ref{fig:nc10k_threshold}. SMPC-based schemes~\cite{khurram2020sfour,weicryptographically} consider the effectiveness of the scheme in terms of high recall. %The precision is supposed to be achieved based on the effectiveness of the blocking scheme. In terms of recall, 
SFour~\cite{khurram2020sfour} achieved 97.9\% with $\log(n)$ window size, and 98.6\% (best) and secure PRL~\cite{weicryptographically} achieved 99.97\% over the NCVR dataset. Under comparable settings over 10k records, our clustering-based search has 96.1\%, while linear search achieved 98.6\% recall. The drop in recall is due to clustering and homomorphic error. However, our approach achieved perfect precision, which is of greater significance, in the financial context.

\begin{comment}
\begin{multicols}{2}
\begin{table}[H]
\caption{Accuracy comparison (NCVR)}
\label{tab:comp}
\centering
\resizebox{0.48\textwidth}{!}{
\begin{tabular}{|c|c|c|c|c|}
\hline
SFour~\cite{khurram2020sfour}  & SFour (Best) & SPRL~\cite{weicryptographically} & Ours & Ours (Best) \\ \hline
97.9 & 98.6  & 99.97                                 & 96.1              & 98.6        \\ \hline
\end{tabular}}
\end{table}

\begin{table}[H]
\centering
\caption{Performance comparison.}
\label{tab:nc-pc}
\resizebox{0.48\textwidth}{!}{
\begin{tabular}{|c|c|c|c|c|c|}
\hline
Schemes & SFour~\cite{khurram2020sfour} & SPRL~\cite{weicryptographically}  & Ours (L) & Ours (C) \\ \hline
Records & 4000  & 40000 & 33000    & 45000   \\ \hline
\end{tabular}}
\end{table}
\end{multicols}
\end{comment}
%\begin{comment}
\begin{table}[h]
\centering
\caption{Performance comparison (in one hour).}
\label{tab:nc-pc}
\begin{tabular}{|c|c|c|c|c|c|}
\hline
Schemes & SFour~\cite{khurram2020sfour} & SPRL~\cite{weicryptographically}  & Ours (L) & Ours (C) \\ \hline
Records & 4000  & 40000 & 33000    & 45000   \\ \hline
\end{tabular}
\end{table}
%\end{comment}
\noindent
\textit{Performance comparison:} Our scheme's performance cannot be directly compared to existing solutions~\cite{essex2019secure,khurram2020sfour,weicryptographically} due to its focus on unbalanced privacy-preserving fuzzy search in contrast to private record linkage. Our solution aligns with regulations that prohibit the sharing of matching records with both parties. The search capability of our scheme allows for finding a match within a few seconds (or minutes), with longer search times limited to data distributions with large clusters. Our scheme requires only 175 KB of ciphertext to send a matching response per column. Table~\ref{tab:nc-pc} compares our work in similar settings for the number of comparisons possible in an hour. It is possible to perform a private fuzzy search between two datasets with 33,000 records, using our algorithm employing linear search. With clustering, it can be scaled to 45,000. We refer to Appendix~\ref{sec-results} for more results.

\section{Conclusion}
%\noindent\textcolor{red}{Cormode: In conclusion, can we say we have a system that achieves sufficient accuracy to make banks and regulators happy?\\ Can we recommend appropriate settings for the system?
%In future work, we can analyse the scenario of a malicious entity (for verifiable computation). The client can be the malicious one, not the server.}\\
%The secure sharing and processing of data across borders has paramount importance in the context of the increasingly interconnected global landscape of international banking. 

This paper explored the intersection of privacy regulations and the increasing requirement for swift data access to strengthen financial security measures. The proposed privacy-preserving fuzzy name matching scheme can be used for approximate string searching in encrypted settings. It preserves the privacy of the query, including the private dataset of the responding organisation. The proposed scheme utilised MinHash signatures with CKKS encryption, which has been proven semantically secure in semi-honest settings. Longer encoding provided better accuracy, but increased communication and computational overheads. Clustering is integrated to reduce the search space and enhance performance. It also reduced the search time, as the results are returned column-wise. While clustering maintains precision, it does bring a drop in recall. 
This gives an opportunity for future work to minimize this recall drop. 
%In future work, there is a scope to improve the recall, without increasing the overheads.

%without revealing either from query  By utilising emerging PETs, specifically homomorphic encryption, we have introduced a novel approach to effectively matching customer names across datasets, while considering the presence of typographical errors and variances in the names.

\bibliographystyle{splncs04}
\bibliography{main}

%\newpage
\appendix

\section{Homomorphic Encryption (HE)}
\label{sec:he}

Formally, let the quotient ring be defined as \( R_Q = \mathbb{Z}_Q[X^N + 1] \), where \( Q \) represents a substantially large modulus integer and \( N \) signifies a degree which is a power of two. This is the polynomial degree over which the CKKS scheme operates. In this context, $Q$ undergoes decomposition into the product of smaller, pairwise co-prime moduli, represented as $Q = \prod_{i=0}^{L} q_i $, where each $q_i$ is one of these smaller moduli. Consequently, this allows for a polynomial $a$ to be  represented in the Residue Number System as $a = \left( [a]_{q_0}, \ldots, [a]_{q_i} \right) \in \prod_{i=0}^{L} R_{q_i}$, where each component is defined as: $[a]_{q_i} = a_0 + a_1X + \ldots + a_{N-1}X^{N-1} \in R_{q_i}$.

\noindent
\textit{Key Generation:}
The secret key $sk$ is constructed by selecting a key $s$ sampled from a distribution $\chi_{\text{key}}$ over $R$. The public key is then formed as a pair $pk = (b, a) \in R_{Q}^2P$, where  $b$ is calculated using the equation $b = -a \cdot s + e$, and $e$ represents the error polynomial from a Gaussian distribution.

\noindent
\textit{Encryption and Decryption:}
To encode a vector of up to $N/2$ real numbers, a plaintext polynomial $m$ with $N$ coefficients modulo $Q$ is used. The plaintext message is then encrypted to produce a ciphertext $ct = (ct_0, ct_1) \in R_{Q}^2P$. The decryption process retrieves the original message as $m' = ct_0 + ct_1 \cdot s \mod Q$, which approximates  the original message, represented as $m' \approx m $. The error introduced during encryption is small and controlled, ensuring that the decrypted message remains a close approximation to the original plaintext. %Homomorphic operations such as addition and multiplication on ciphertexts are described as follows:

\noindent
\textit{Addition ($\operatorname{CKKSAdd}$):}
Given two ciphertexts $ct = (c_0, c_1)$ and $ct' = (c'_0, c'_1)$, the homomorphic addition is $ct_\text{add} = ct + ct' = (c_0 + c'_0, c_1 + c'_1)$. This operation produces a new ciphertext that encrypts the sum of the underlying plaintexts, maintaining the property $m + m' \mod Q$, where $m$ and $m'$ are the encoded plaintexts corresponding to $ct$ and $ct'$ respectively.

\noindent
\textit{Addition by constant ($\operatorname{CKKSAdd_{const}})$:} When a ciphertext $ct = (c_0, c_1)$ is added with a constant (plaintext), it becomes $(c_{0}+m', c_{1})\mod Q$, where the constant is first encoded to $m'$.

\noindent
\textit{Multiplication ($\operatorname{CKKSMult}$):}
The homomorphic multiplication of given ciphertexts $ct = (c_0, c_1)$ and $ct' = (c'_0, c'_1)$ for messages $m$ and $m'$ involves a component-wise product of the ciphertexts, followed by a relinearization step to reduce the degree of the resulting ciphertext as $(c_{\times,0}, c_{\times,1}, c_{\times,2}) = (c_0 \cdot c'_0, c_0 \cdot c'_1 + c_1 \cdot c'_0, c_1 \cdot c'_1).$ 
This is a tensoring operation, where the ciphertext degree is increased from 2 to 3~\cite{agrawal2023high}. The relinearization process, using key-switching, transforms the 3-degree ciphertext back to a 2-degree ciphertext and the final ciphertext $ct_\text{mult}$ satisfies $c_{\times,0} + c_{\times,1} \cdot s + c_{\times,2} \cdot s^2 \approx m \cdot m' \mod Q$, however, it is associated with the squared scaling factor which is later reduced back by the rescaling procedure. 

\noindent
\textit{Multiplication by constant ($\operatorname{CKKSMult_{const}})$:} When a ciphertext $ct = (c_0, c_1)$ is multiplied by a constant (plaintext), it becomes $(c_{0}.'m, c_{1}.m') \mod Q$, where the constant is first encoded to $m'$.

On top of this, HE supports batching, which packs multiple messages in one ciphertext, thus enabling one operation to act over all the messages. It improves performance both in terms of computation and communication.

\begin{algorithm}[!t]
\caption{MinHash signature generation}
\label{alg:lsh}
\KwData{Input Name $N$,\\ 
\setlength\parindent{27pt} Shingle Size $S$, \\
\setlength\parindent{27pt} Number of Permutations $P$} 
\KwResult{MinHash signature $\vec{N}$ of length $P$}
\Begin{
    \textbf{Function} $\operatorname{GenerateShingles}$ ($N$)\;
    \Begin{
        shingles $\leftarrow \emptyset$; \Comment{sets of characters}\;
        \For{$i \leftarrow 0$ \textbf{to} $\operatorname{length}(N) - S$}{
            shingle $\leftarrow \operatorname{SubString}$ ($N$, $i$, $i+S$)\;
            shingles.$\operatorname{add}$(shingle)\;
        }
        \Return shingles\;
    }  
    
    %\textbf{function} HashShingle($shingle$)\;
    %\Begin{
    %    \Return SHA-256($shingle$)\;
    %}
    shingles $\leftarrow \operatorname{GenerateShingles} $($N$)\;
    hashes $\leftarrow \emptyset$\;
    \ForAll{\textrm{shingle} $\in$ shingles}{
        hash $\leftarrow \operatorname{HashFunction}$ (shingle)\;
        hashes.$\operatorname{add}$(hash)\;
    }
    \For{$i \leftarrow 1$ \textbf{to} $P$}{
        $\vec{N}[i] \leftarrow \infty$\;
        \ForAll{\textrm{hash} $\in$ hashes}{
            PermutedHash $\leftarrow$ $i$ random permutation functions of hash\;
            $\vec{N}[i] \leftarrow$ $\operatorname{min}$($\vec{N}[i]$, PermutedHash)\;
        }
    }
    \Return $\vec{N}$\;
    }
\end{algorithm}

\begin{algorithm}[!t]
%\resizebox{0.48\textwidth}{!}{
\caption{Offline Data Preparation - Clustering}
\label{alg:clustering}
\begin{minipage}{\linewidth}
\KwData{MinHash signatures $\vec{N_B}$}
\KwResult{Clustered matrix $(C)$ and Centroids $(\vec{C_k})$}

\textbf{Function} $\operatorname{Clustering}$ ($\hat{N_B}, k, iterations$)\;
\Begin{
    Initialize random centroids $\vec{C_k} = \{\vec{c_1}, \vec{c_2}, \ldots, \vec{c_k}\}$\;

    $\hat{N_{B}} \gets \frac{\vec{N_B}}{|\vec{N_B}|}$\;
    $\hat{N_{Bs}} \gets \operatorname{StandardScaler}(\hat{N_B})$\;
    \Repeat{iterations}{
        Assign each $\hat{N_{Bs}}$ to the nearest centroid based on the cosine similarity measure\;
        Update clusters ($C$) and centroids ($\vec{C_k}$) using KMeans\;
    }

    $max\_size \gets $ maximum number of elements in any cluster of $C$; \Comment{Need to pad remaining clusters to this size}\;
    \For{each \textrm{cluster} in $C$}{
        Pad $max\_size - \operatorname{len}(\textrm{cluster})$ dummy elements.
    }
    
    \Return clustered matrix $(C)$ and list of centroids $(\vec{C_k})$;% = \{\vec{c_1}, \vec{c_2}, \ldots, \vec{c_k}\}$;
}

\Begin{
    $B$ chooses the number of clusters $k$ close to $\sqrt(|\vec{N_B}|)$\;
    Set the maximum number of repeats $iterations$\;
    $C, \vec{C_k} \gets  \operatorname{Clustering}$($\vec{N_B}, k, iterations$)\;
}

\end{minipage}
%}
\end{algorithm}

\begin{algorithm}[!t]
%\resizebox{0.48\textwidth}{!}{
\caption{CKKS Dot Product}
\label{alg:dotp}
\begin{minipage}{\linewidth}
\KwData{ciphertext message 1 $ct_{m1}$, %\\ \setlength\parindent{27pt} 
ciphertext message 2 $ct_{m2}$/plaintext message $pt_{m2}$}
\KwResult{Encrypted Dot Product $dotp$}
\smallskip

\textbf{Function} $\operatorname{DotProduct}$ ($ct_{m1}$, $pt_{m2}$)\;
\Begin{
    size $\gets $ length of the ciphertext vector\;
    $\operatorname{vector} \langle\text{Ciphertext} \rangle$ dotp\;
    %\# parallelised for loop\;
    \For{$i \gets 0$ \textbf{to} $\textrm{size} - 1$}{
        $m \gets \operatorname{CKKSEncode}(pt_{m2}[i])$\;
        $ct_{mult} \gets \operatorname{CKKSMult_{const}}(ct_{m1}[i], m)$\;
        $dotp \gets \operatorname{CKKSAdd}(dotp, ct_{mult})$
    }
    \Return $dotp$;
}

\textbf{Function} $\operatorname{DotProduct}$ ($ct_{m1}$, $ct_{m2}$)\;
\Begin{
    size $\gets $ length of the ciphertext vector\;
    $\operatorname{vector} \langle\text{Ciphertext} \rangle$ dotp\;
    %\# parallelised for loop\;
    \For{$i \gets 0$ \textbf{to} $\textrm{size} - 1$}{
        $ct_{mult} \gets \operatorname{CKKSMult}(ct_{m1}[i], ct_{m2}[i])$\;
        $dotp \gets \operatorname{CKKSAdd}(dotp, ct_{mult})$
    }
    \Return $dotp$;
}

\end{minipage}
%}
\end{algorithm}

\section{Algorithms}\label{sec:algorithms}
This section presents the complete procedure for MinHash signature generation in Algorithm~\ref{alg:lsh}, clustering in Algorithm~\ref{alg:clustering} and CKKS dot product in Algorithm~\ref{alg:dotp}. These algorithms have been used in the proposed scheme (Section~\ref{V.B}) for performing different operations, including dataset encoding, clustering, and encrypted cosine similarity.

\section{Theoretical Analysis}~\label{sec:theoanal}
%\noindent\textcolor{red}{Cormode: What are the properties of the encryption mechanism? How can it fail?}\\
%\noindent\textcolor{blue}{Cormode: Can we consider any post processing to reduce false positives?}\\

%Security can be defined in terms of indistinguishability according to assumptions made about an adversary's capabilities~\cite{baritel2020formal}. IND-CPA ensures data confidentiality, \textit{i.e.}, the adversary cannot retrieve the plaintext (here, encoded LSH query) without knowing the secret key (\textit{sk}). Thus, if the proposed scheme is IND-CPA secure, it suffices the requirement. IND-CPA relies on the idea that the adversary cannot distinguish between ciphertexts corresponding to different plaintexts.

%\textit{Q} initiates the protocol, which first uses Locality-sensitive hashing (LSH) to encode the query, and then CKKS FHE to encrypt the encoded query.

This section presents the theoretical analysis of the proposed scheme. First, we give definitions for semantic security under chosen plaintext attack, semi-honest setting and secure matching. Utilising that, we perform the privacy analysis. Further, the correctness analysis discusses the error bounds with homomorphic operations.

%\subsection{Definitions}

\begin{definition}[Semantic Security Under Chosen Plaintext Attack]\label{def-sem-sec}
%\noindent \textit{Definition 2 (Semantic Security Under Chosen Plaintext Attack):} 
An encryption scheme $E$ is \emph{semantically secure under a chosen plaintext attack} if, for any efficient (polynomial-time) adversary $\mathcal{A}$ that is allowed to choose plaintexts and observe their corresponding ciphertexts, $\mathcal{A}$ cannot distinguish between the encryption of two equal-length messages. Formally, for all pairs of plaintexts $N_A$ and $N_B$, \[
\operatorname{Pr}\left[\mathcal{A}\left(\operatorname{Enc}\left(N_A\right)\right)=1\right] \approx \operatorname{Pr}\left[\mathcal{A}\left(\operatorname{Enc}\left(N_B\right)\right)=1\right].\] Even with the ability to choose plaintexts and after receiving the corresponding ciphertexts, $\mathcal{A}$ should not be able to gain any information about the original plaintext or distinguish between them. This is fundamental to ensure the confidentiality of encrypted messages.
\end{definition}

%\begin{comment}
\begin{definition}[Semi-Honest Setting]
%\noindent \textit{Definition 3 (Semi-Honest Setting):} 
Under the \emph{semi-honest setting}, the 
two organisations \textit{A} and \textit{B} follow the scheme faithfully, but may try to infer about the other's data. 
%honestly, but curious to learn/infer from the other's encrypted message. 
Suppose \textit{A} and \textit{B} have inputs  $N_A$ and $N_B$, respectively. 
Then, the view of \textit{A} for the execution of the scheme over the combined input ($N = N_A, N_B$), is $\operatorname{View_{Org_A}}(N)=\left(N_A, O_A, I_A\right)$, where $N_A$ is input, $O_A$ is all the messages generated, and $I_A$ is all the messages received during the execution of the scheme by \textit{A}. The final output of the scheme for \textit{A} is denoted as $\operatorname{Out_{Org_A}}$. Similarly, the view of \textit{B} is $\operatorname{View_{Org_B}}(N)=\left(N_B, O_B, I_B\right)$ and the output is denoted as $\operatorname{Out_{Org_B}}$.
\end{definition}
%\end{comment}

\begin{definition}[Secure Matching]\label{def-sec-match}
%\noindent \textit{Definition 4 (Secure Matching):} 
The scheme securely evaluates the matching function $(S)$ in the semi-honest threat model if there exist probabilistic polynomial-time algorithms $P_A^* \text { and } P_B^*$, such that
\[\left[P_A^*\left(N_A, M(N)\right), M(N)\right] \stackrel{c}{\equiv}\left[\operatorname{View_{Org_A}}(N), \operatorname{Out_{Org_B}}(N)\right]\]
and
\[\left[P_B^*\left(N_B, M(N)\right), M(N)\right] \stackrel{c}{\equiv}\left[\operatorname{View_{Org_B}}(N), \operatorname{Out_{Org_A}}(N)\right].\]% where $[]$ denotes fully homomorphic (CKKS) encryption. 
\end{definition}

\subsection{Privacy Analysis}

As discussed in the threat model, we consider a semi-honest (\textit{honest-but-curious}) threat model. It means both organisations, querying (\textit{A}) and responding (\textit{B}), execute the secure matching (Definition \ref{def-sec-match}), but are curious to learn/infer facts from the other's messages.

\begin{comment}
\begin{lemma}[Semantic Security of the Encryption Scheme]
    The public-key encryption scheme is semantically secure under a chosen plaintext attack, and ensures data confidentiality.
\end{lemma}

\textit{Proof.} See Appendix~\ref{sec-th1}.
\qed
\end{comment}

\begin{theorem}[Querying Organisation \textit{A}'s Privacy] 
Let \textit{A}'s input to the scheme be $(a, pk, sk)$ and \textit{B}'s input be $(b)$, and their combined input be $x$. $\operatorname{View_{Org_B}}$ represents the view of \textit{B} during the execution and $\operatorname{Out_{Org_A}}$ is the output of \textit{A}. 
Then there exists a probabilistic polynomial-time algorithm $P_B^*$, such that, 
\[\left[P_B^*\left(b, \perp\right), F(x)\right] \stackrel{c}{\equiv}\left[\operatorname{View_{Org_B}}(x), \operatorname{Out_{Org_A}}(x)\right],\]
where $\perp$ denotes no output and $F$ denotes the functions defined in Algorithm~\ref{alg:scheme1}.
\end{theorem}

\textit{Proof.} Organisation \textit{A}'s privacy can be proved simply because \textit{B} receives only a fixed number of ciphertexts from \textit{A}. Ciphertexts are computationally indistinguishable~\cite{baritel2020formal}, because of the semantic security of the encryption scheme (Definition~\ref{def-sem-sec}). To simulate \textit{B}'s view, $P_B^*$ samples a random MinHash signature of the agreed encoding size, and encrypts it using \textit{A}'s public key (Step 2 in Section~\ref{V.B}). Then, it generates an array of size equal to the number of centroids, with 0's and 1 (at any random position), followed by encryption using \textit{A}'s public key (Step 5 in Section~\ref{V.B}). The remaining steps can be simulated at \textit{B} over its inputs.
\qed

\begin{theorem}[Responding Organisation \textit{B}'s Privacy] Let $\operatorname{View_{Org_A}}$ represent the view of \textit{A} during the execution and all other inputs as defined above. Then, there exists a probabilistic polynomial-time algorithm $P_A^*$, such that,
\[\left[P_A^*\left(a, F(x)\right), \perp\right] \stackrel{c}{\equiv}\left[\operatorname{View_{Org_A}}(x), \perp\right].\]
\end{theorem}

\textit{Proof.} Organisation \textit{B}'s privacy relies on the randomness of the result. To simulate \textit{B}'s view, \textit{A} needs to generate random samples to match the outputs generated by Step 8 in Algorithm~\ref{alg:scheme1}. The original outputs returned by \textit{B} are the scores, \textit{i.e.}, $r * (E(\text{cos\_score}) - \tau)$, where $\text{cos\_score}$ is the cosine similarity between the encoding of \textit{A}'s input and encodings of \textit{B}, $\tau$ is the threshold defined for cosine similarity match, and $r$ is a random number ($r \in \mathbb{Z}_p^*$). Cosine similarity ranges between 0 and 1. Subtracting $\tau$ rescales it between $-\tau$ and $1-\tau$. Multiplying $r$ with $\text{cos\_score} - \tau$ results in either a positive or negative number, which depends on the sign value of $\text{cos\_score} - \tau$. Since the returned score is random, due to $r$, it does not leak anything about actual cosine similarity, other than revealing whether there is a potential match or not.
\qed

\subsection{Correctness}
The correctness of our proposed scheme depends on the cosine similarity of encoded names (MinHash signatures). To calculate cosine similarity, we implement a CKKS dot (inner) product between normalised ciphertext-plaintext and ciphertext-ciphertext encoded names. Thus, we prove the correctness of the dot product under both scenarios. First, we describe our assumptions based on the arithmetic of approximate numbers in CKKS~\cite{cheon2017homomorphic,hu2024faster}.

\begin{assumption} A ciphertext $\left(ct \in \mathcal{R}_{q_{\ell}}^2, \ell, \nu, B\right)$ is a valid encryption of $m \in \mathcal{S}$ if $\|m\|_{\infty}^{\mathrm{can }} \leq \nu$ and $\langle ct, s k\rangle=m+e\left(\bmod q_{\ell}\right)$ for some polynomial $e \in \mathcal{S}$ with $\|e\|_{\infty}^{\mathrm{can }} \leq B$.
\end{assumption}

\begin{assumption} The noise due to encoding and encryption is bounded by $B_\mathrm{clean} = 8 \sqrt{2} \sigma N+$ $6 \sigma \sqrt{N}+16 \sigma \sqrt{h N}$.
\end{assumption}

\begin{assumption}
The noise due to rescaling is bounded by $B_\mathrm{scale} = \sqrt{N / 3} \cdot(3+8 \sqrt{h})$.
\end{assumption}

\begin{assumption}
For addition and multiplication by plaintext (constant) $a$, \\ $\left(ct_{\mathrm{add}}, \ell, \nu+\|a\|_{\infty}^{\mathrm{can }}, B\right)$ and $\left(ct_{\mathrm{mult}}, \ell,\|a\|_{\infty}^{\mathrm{can }} \cdot \nu,\|a\|_{\infty}^{\mathrm{can }} \cdot B\right)$ are valid encryptions of $m+a$ and $am$, respectively.%, where $\|a \cdot e\|_{\infty}^{\text {can }} \leq\|a\|_{\infty}^{\text {can }} \cdot B$.
\end{assumption}

\begin{assumption}
For addition and multiplication between any two ciphertexts, $ct_{1}$ and $ct_{2}$, $\left(ct_{\mathrm{add}}, \ell, \nu_1+\nu_2, B_1+B_2\right)$ is valid encryption of $m_1+m_2$ and \\
$\left(ct_{\mathrm{mult}}, \ell, \nu_1 \nu_2, \nu_1 B_2+\nu_2 B_1+B_1 B_2+B_{\mathrm{mult }}(\ell)\right)$ for $m_1 \cdot m_2$, where $B_{\mathrm{ks}}=8 \sigma N / \sqrt{3}$ and $B_{\mathrm{mult }}(\ell)=P^{-1} \cdot q_{\ell} \cdot B_{\mathrm{ks}}+B_{\mathrm{scale }}$. %$v_1$ and $v_2$ are numbers greater than $\|m_1\|_{\infty}^{c a n},\left\|m_2\right\|_{\infty}^{c a n}$, respectively. $B_{1}$ and $B_{2}$ are upper bounds of the noise of $c_{1}$ and $c_{2}$, respectively.
\end{assumption}

\begin{theorem} Let $m_{1}=P_N(v_{1}), m_{2}=P_N\left(v_2\right)$ be the polynomial representations of $\frac{N}{2}$-dimensional vectors $v_1, v_2$, respectively. The ciphertext $ct_{1} \in R_{q_l, N}$ is the encryption of $m_{1}$, while $m_2$ is the encoding of a plaintext (constant) vector $v_2 = [a_{1}, a_{2}, \ldots, a_{\frac{N}{2}}]$. Algorithm~\ref{alg:dotp} takes the input of ciphertext and plaintext, then returns the ciphertext of a polynomial $P_N\left(\left\langle v_1,  v_2\right\rangle,\left\langle v_1,  v_2\right\rangle, \ldots,\left\langle v_1,  v_2\right\rangle\right)$ with a noise bounded by $B_{\mathrm{dotp }} \leq \frac{N}{2}\cdot\|a\|_{\infty}^{\mathrm{can }} \cdot B$.
\end{theorem}

\textit{Proof.} Consider two vectors of size $\frac{N}{2}$, with an encrypted vector $\vec{ct_1} = (ct_{11}, ct_{12}, \ldots, ct_{1\frac{N}{2}})$ and a plaintext vector $(a_1, a_2, \ldots, a_\frac{N}{2})$, the dot product of these two vectors is computed as $\sum_{i=1}^{\frac{N}{2}} (ct_{1i} \cdot a_i)$.

Based on Assumptions 1, 2, and 4, and a polynomial $e \in \mathcal{R}$ such that $\langle ct, s k\rangle=m+e\left(\bmod q_{\ell}\right)$ and $\|e\|_{\infty}^{\text {can }} \leq B$. It is obvious that $\left\langle ct_{\mathrm{m}}, s k\right\rangle=$ $a \cdot(m+e)=a m+a e\left(\bmod q_{\ell}\right)$ and $\|a \cdot e\|_{\infty}^{\text {can }} \leq\|a\|_{\infty}^{\text {can }} \cdot B$.

Based on Assumption 5, the addition of ciphertexts is bounded by the sum of the upper bounds of their respective noises. Therefore, the noise is bounded by $B_{\text {dotp }}$, \textit{i.e.}, $\frac{N}{2}\cdot\|a \cdot e\|_{\infty}^{\text {can }} \leq \frac{N}{2} \cdot \|a\|_{\infty}^{\text {can }} \cdot B$.
\qed

\begin{theorem} Let $m_{1}=P_N(v_1), m_{2}=P_N\left( v_2\right)$ be the polynomial representations of $\frac{N}{2}$-dimensional vectors $ v_1,  v_2$, respectively. The ciphertexts $ct_{1}, ct_{2} \in R_{q_l, N}$ are the encryptions of $m_1, m_2$, respectively. Algorithm~\ref{alg:dotp} takes the input of two ciphertexts, then returns the ciphertext of a polynomial $P_N\left(\left\langle v_1,  v_2\right\rangle,\left\langle v_1,  v_2\right\rangle, \ldots,\left\langle v_1,  v_2\right\rangle\right)$ with a noise bounded by $B_{\mathrm{dotp }} \leq \frac{N}{2} B_\mathrm{m u}+\frac{N}{2} B_\mathrm{m u l t}(l)$, where $B_\mathrm{m u}=\left(v_1 B_2+v_2 B_1+B_1 B_2\right)$.
\end{theorem}

\textit{Proof.} This involves the summation of ciphertexts produced by the multiplication of ciphertexts. Given ciphertexts $ct_1$ and $ct_2$ as  $\left(m_1 + e_{10}, e_{11}\right) \bmod q_l$ and $\left(m_2+e_{20}, e_{21}\right) \bmod q_l$, where $e_{10}+e_{11} s=$ $e_1 \bmod q_l$ and $e_{20}+e_{21} s=e_2 \bmod q_l$. With the evaluation key as $\left(P s^2+e_0^{\prime}, e_1^{\prime}\right)$, where $e_0^{\prime}+e_1^{\prime} s=e^{\prime} \bmod P q_l$. Then, the noise for the ciphertext is $m_1 e_2+m_2 e_1 + e_1 e_2 + \frac{e_{11} e_{21} e^{\prime}}{P}+e_\mathrm{s c a l e, 10} \bmod q_l$ bounded by $B_{m u}$.

Based on Assumptions 1, 2, and 5, the noise for the multiplication of two ciphertexts is bounded by $\nu_1 B_2+\nu_2 B_1+B_1 B_2+B_{\text {mult }}(\ell)$, where $B_{\mathrm{mult }}(\ell)=P^{-1} \cdot q_{\ell} \cdot B_{\mathrm{ks}}+B_{\text {scale }}$ is the noise induced due to relinearization. $v_1$ and $v_2$ are numbers greater than $\|m_1\|_{\infty}^\mathrm{c a n},\left\|m_2\right\|_{\infty}^\mathrm{c a n}$, respectively. $B_{1}$ and $B_{2}$ are upper bounds of the noise of $ct_{1}$ and $ct_{2}$, respectively. $\sigma$ is the standard variation of the original noise.

Based on Assumption 5, the addition of ciphertexts is bounded by the sum of the upper bounds of their respective noises. Therefore, the noise is bounded by $B_{\text {dotp }}$, \textit{i.e.}, $\frac{N}{2} B_\mathrm{m u}+\frac{N}{2} B_\mathrm{m u l t}(l)$.
\qed

\section{Datasets}\label{sec:datasets}
The first dataset is derived from the North Carolina Voter Registration (NCVR) Statistics~\cite{NCSBE}, which North Carolina has been maintaining since 2005 to track both active and inactive voters and provide temporal snapshots of voter data within the system. The snapshots dated January 1, 2014, and January 1, 2017, are selected for the study. North Carolina identification number (NCID) field is used as the identifier for establishing the ground truth of a matching record. The concatenation of the first name, middle name, last name, and name suffix was used to generate the full name as registered. %Additionally, the age data were converted to the year of birth to experiment with its impact on the rate of false positives. From each of the two selected datasets, the subsets are extracted with incrementally increasing sizes: 10,000, 100,000, and 1,000,000 records, respectively. 
These records are chosen based on the lowest NCID values to ensure no duplication. Out of 1,000,000, 912,989 records share the same NCIDs.

The second dataset integrates two library catalogue datasets: (1) shadow library book downloads sourced from one of LibGen's mirror sites~\cite{bodo2020shadow}, and (2) a collection of books from bookdepository.com~\cite{simakis_2020}. This dataset encompasses the metadata of books, including ISBN, title, and author. We employed the ISBN to establish the ground truth and title for fuzzy matching. We applied a filter to select records where the titles have at least a $90\%$ similarity based on the Levenshtein distance between the matching pairs, similar to related work~\cite{khurram2020sfour}, and we selected 18,636 matching records across the two datasets. 
Despite not being directly related to our use case, we use this dataset because it contains longer names.

Utilising data sourced from the US Census~\cite{us_census}, the third dataset is constructed, featuring the most frequently occurring given and family names. We generated random pairs of given and family names, each assigned a unique ID number. For every pair, we then applied random modifications for Levenshtein distances that varied from 1 to 5 to introduce five fuzzy variants to each name pair in the dataset. We study this dataset to show the effectiveness of the scheme to varying distances.

\section{Additional Results}\label{sec-results}

\begin{figure}
    \centering
    %\resizebox{0.33\textwidth}{0.33\textwidth}{%
    \begin{tikzpicture}[thick, scale=0.5, every node/.style={transform shape}]
    \begin{axis}[
        %label style={font=\huge},         ticklabel style = {font=\Large},
        legend columns=2, 
        legend style={at={(0.35,0.2)},anchor=north,draw={none}},
        xlabel= Levenshtein Distance (LD),
        ylabel= Percentage,
        ymin=0, ymax=1,
        xticklabels={0,1,2,3,4,5},
        ytick={0,.20,.40,.60,.80,1},
        grid=both,
        xtick=data,
        xtick distance=1,
        table/x expr=\coordindex,
        ]

    \addplot[color=black,mark=+] table [x=LD, y=Accuracy, col sep=comma] {results/us10k.csv};
    \addlegendentry{Accuracy}
    \addplot[color=red,mark=x] table [x=LD, y=Precision, col sep=comma] {results/us10k.csv};
    \addlegendentry{Precision}
    \addplot[color=cyan,mark=*] table [x=LD, y=Recall, col sep=comma] {results/us10k.csv};
    \addlegendentry{Recall}
    \addplot[color=blue,mark=o] table [x=LD, y=F1-Score, col sep=comma] {results/us10k.csv};
    \addlegendentry{F1-Score}
    
    \end{axis}
    \end{tikzpicture}%}
    \caption{Sensitivity analysis on US Census dataset.}
    \label{fig:us10k}
\end{figure}
Fig.~\ref{fig:us10k} studies the sensitivity of the proposed scheme based on varying Levenshtein Distance (LD) over the US Census dataset. It can be observed that fuzzy names with LD 0 and 1 are getting searched with almost 100\% precision and recall. Recall drops to 70\% with LD2 and up to 10\% with LD5 while maintaining the precision.

\begin{table}[!t]
\centering
\caption{Performance Scaling Evaluation.}
\label{tab:nc-sc}
\begin{tabular}{|c|c|c|c|c|}
\hline
No of & first round & time/col & memory & comm/col             \\ \hline
queries        & secs        & secs     & GB     & KB                   \\ \hline
1       & 0.72        & 0.26     & 6.63    & \multirow{4}{*}{175} \\ \cline{1-4}
10      & 1.4         & 0.26     & 6.63   &                      \\ \cline{1-4}
100     & 1.53        & 0.26     & 6.63   &                      \\ \cline{1-4}
1000    & 1.98        & 0.26     & 6.63   &                      \\ \hline
\end{tabular}
\end{table}

Table \ref{tab:nc-sc} logs the evaluation results with varying numbers of queries. The result is reported over the NCVR dataset with 10k, and the number of clusters is set as 50 (308 columns).
%However, this can be generalised across datasets.
The number of queries is varied from 1 to 1000. Batching with more than 1000 queries failed due to higher memory requirements in the current simulation environment. It can be observed that, increasing the number of queries causes a marginal increase in first-round computation time. Other computation and communication costs are constant. Thus, it does not impact overall search time. %In summary, the proposed unbalanced fuzzy secure and private search algorithm is scalable.

\begin{table}[h]
\centering
\caption{Computation and Communication costs.}
\label{tab:nc-he}
\begin{tabular}{|c|c|c|c|}
\hline
         & Ciphertext & Time & Configuration                                          \\ \hline
AHE~\cite{essex2019secure}      & 76 GB      & 1.8 H         & Intel Xeon CPU E5-2697A \\ \hline
Ours (L) & 63 GB      & 0.6 H         & \multirow{2}{*}{Intel Xeon CPU E5-4650 v4}          \\ \cline{1-3}
Ours (C) & 1 GB       & 0.45 H        &                                                        \\ \hline
\end{tabular}
\end{table}

Compared with additive homomorphic encryption-based solution~\cite{essex2019secure}, our algorithm achieved improved results both in terms of computation and communication. Table~\ref{tab:nc-he} reports the result in terms of generated ciphertext and time taken for matching 20,000 fuzzy records between two parties. AHE-based solutions incur higher communication costs and take at least thrice the time to search compared to our algorithm. In summary, the proposed algorithm is more secure and scalable for privacy-preserving fuzzy name matching.

\end{document}